\documentclass[%
reprint,
superscriptaddress,
showpacs,
nofootinbib,
 amsmath,amssymb,
 aps,
]{revtex4-2}

\usepackage[utf8]{inputenc}

\usepackage{setspace}
\usepackage[utf8]{inputenc}
\usepackage[english]{babel} 
\usepackage{graphicx}
\usepackage{physics}
\usepackage{booktabs}
\usepackage{multirow}
\usepackage[colorlinks=true,linkcolor=blue,citecolor=blue]{hyperref}

\usepackage{tikz}
\usetikzlibrary{arrows,shapes}
\usetikzlibrary{trees}
\usetikzlibrary{matrix,arrows} 				
\usetikzlibrary{positioning}				
\usetikzlibrary{calc,through}				
\usetikzlibrary{decorations.pathreplacing}  
\usepackage{pgffor}							
\usetikzlibrary{decorations.pathmorphing}	
\usetikzlibrary{decorations.markings}
\tikzset{
	vector/.style={decorate, decoration={snake}, draw},
	provector/.style={decorate, decoration={snake,amplitude=2.5pt}, draw},
	antivector/.style={decorate, decoration={snake,amplitude=-2.5pt}, draw},
	fermion/.style={draw=black, postaction={decorate},
	decoration={markings,mark=at position .55 with {\arrow[draw=black]{>}}}},
	fermionbar/.style={draw=black, postaction={decorate},
	decoration={markings,mark=at position .55 with {\arrow[draw=black]{<}}}},
	fermionnoarrow/.style={draw=black},
	gluon/.style={decorate, draw=black,
	decoration={coil,amplitude=4pt, segment length=5pt}},
	scalar/.style={dashed,draw=black, postaction={decorate},
	decoration={markings,mark=at position .55 with {\arrow[draw=black]{>}}}},
	scalarbar/.style={dashed,draw=black, postaction={decorate},
	decoration={markings,mark=at position .55 with {\arrow[draw=black]{<}}}},
	scalarnoarrow/.style={dashed,draw=black},
	electron/.style={draw=black, postaction={decorate},
	decoration={markings,mark=at position .55 with {\arrow[draw=black]{>}}}},
	bigvector/.style={decorate, decoration={snake,amplitude=4pt}, draw},
}

\usepackage{ulem}

\begin{document}


\vspace{-6pt}
\title{
	Model-independent test of T violation in neutrino oscillations
}

\author{Thomas Schwetz}%
\affiliation{%
	Institut für Astroteilchenphysik, Karlsruher Institut für Technologie (KIT), 
	76131 Karlsruhe, Germany
}%
\author{Alejandro Segarra}%
\affiliation{%
	Institut für Theoretische Teilchenphysik, Karlsruher Institut für Technologie (KIT), 
	76131 Karlsruhe, Germany
}%

\date{\today}

\begin{abstract}
  We propose a method to establish
  time reversal symmetry violation at future neutrino oscillation
  experiments in a largely model-independent way. We introduce a general
  parametrization of flavour transition probabilities which holds
  under weak assumptions and covers a large class of new-physics
  scenarios. This can be used to search for the presence of T-odd
  components in the transition probabilities by comparing data at
  different baselines but at the same neutrino energies. We show that
  this test can be performed already with experiments at three different baselines and might be feasible with experiments under preparation/consideration.
\end{abstract}


\maketitle


\textbf{Introduction.}
The violation of time reversal (T) and charge-parity (CP) symmetries
are central topics in particle physics. CP violation (CPV) is one of
the necessary conditions to generate a matter-antimatter asymmetry in
the early Universe \cite{Sakharov:1967dj}, and under the well founded
assumption of CPT conservation, CPV is equivalent to T violation
(TV). A particularly active field is the search for CPV in neutrino
oscillations~\cite{Cabibbo:1977nk,Bilenky:1980cx,Barger:1980jm}. Unfortunately,
the experimental signature is rather indirect, and it is not possible
to construct model-independent CP-asymmetric observables in neutrino
oscillation experiments. This is related to the fundamental obstacle
that experiments and detectors are made out of matter (and not
antimatter). Moreover, the passage of the neutrino beam through Earth
matter introduces environmental CPV due to
matter effects~\cite{Wolfenstein:1977ue}.

The standard approach to this problem is to perform a model-dependent
fit to data. This involves the assumptions that neutrino production,
detection and propagation is fully understood in terms of Standard Model
(SM) interactions, that neutrino mixing is unitary, and only the three
SM neutrino flavours exist. In this case oscillation physics can be
parametrized in terms of a unitary $3\times3$ lepton-mixing
matrix~\cite{Pontecorvo:1957cp,Maki:1962mu} and two neutrino
mass-squared differences. CPV is then described by a complex phase
$\delta$ in the mixing matrix~\cite{Cabibbo:1977nk,Kobayashi:1973fv} which can be fitted against
data. ``Observation of CPV'' is considered equivalent to establishing
that $\delta$ is different from 0 and $\pi$ at a certain confidence
level. Within this restricted framework, current data start to
provide first indications of preferred regions for the parameter
$\delta$~\cite{Abe:2019vii,Acero:2019ksn,Esteban:2020cvm,deSalas:2020pgw,Capozzi:2020qhw}.

Large activity is devoted to study the impact of non-standard
scenarios on the search for CPV in neutrino oscillations. Examples are
non-unitary mixing~\cite{FernandezMartinez:2007ms,Escrihuela:2016ube}, non-standard neutrino interactions~\cite{Ge:2016dlx,deGouvea:2015ndi,Denton:2020uda}, or the presence
of sterile neutrinos~\cite{Gandhi:2015xza,Palazzo:2015gja,Berryman:2015nua}. 
In such new-physics scenarios, additional complex phases appear, which can act as new sources for CP and T violation.
Typically one adopts a specific parameterization
of new-physics and again performs a parametric fit in the extended
model. Our aim in this letter is to go a step beyond such approaches
and develop a largely model-independent test, covering a wide class of
non-standard scenarios.
Our approach is based on fundamental principles about TV noted in the seminal paper by N. Cabibbo~\cite{Cabibbo:1977nk}.

\bigskip\textbf{Model-independent description of flavour evolution.}
Here we specify our approach to describe the
neutrino survival and transition probabilities $P_{\alpha\beta}$, with
$\alpha,\beta = e,\mu,\tau$.  $P_{\alpha\beta}$ is the probability for a
neutrino $\nu_\alpha$ produced at the neutrino source to arrive as
$\nu_\beta$ at the detector. We adopt the following assumptions:

$(i)$ Propagation of the three SM neutrino states is described by a
hermitian Hamiltonian $H(E,x)$, which depends on neutrino energy $E$
and in general on the matter density at the position $x$ along the
neutrino path.

$(ii)$ We assume that for the experiments of interest, medium
effects can be described to sufficient
accuracy by a constant matter density which is approximately the same for all considered experiments. This is a good approximation for
experiments with baselines less than several
1000~km~\cite{Miura:2001pi,Yokomakura:2002av}. 
In an accompanying paper [PRD] we show that for the baselines relevant for our test, the effect of non-constant density is negligible.

Assumption $(ii)$ implies that the matter effect does not introduce
environmental TV by itself~\cite{Krastev:1988yu,Akhmedov:2001kd} [PRD]. Therefore, any observation of
TV can be related to fundamental TV of the theory. (General discussions about TV in neutrino oscillations can be found e.g.\ in~\cite{Cabibbo:1977nk,Kuo:1987km,Toshev:1989vz,Arafune:1996bt,Parke:2000hu,Xing:2013uxa,Petcov:2018zka,Bernabeu:2018twl,Bernabeu:2018use,Bernabeu:2019npc}.) Furthermore, the
Hamiltonian becomes position independent and we can diagonalize it as
$H(E) = W \lambda W^\dagger$, with $W$ being a unitary matrix and
$\lambda = (\lambda_i)$ is a diagonal matrix of the real eigenvalues
$\lambda_1,\lambda_2,\lambda_3$ of $H$. Both $W$ and $\lambda$ depend
on the neutrino energy. Note that we allow for arbitrary non-standard
matter effects. In general, $W$ and $\lambda$ will be different for
neutrinos and antineutrinos.

$(iii)$ We allow for arbitrary (non-unitary) mixing of the energy
eigenstates $\nu_i$ with the flavour states $\nu_\alpha$ relevant for
detection and production,
\begin{equation}\label{eq:mixing}
  |\nu_\alpha\rangle = \sum_{i=1}^3 N_{\alpha i}^\text{prod,det} |\nu_i\rangle \,.
\end{equation}
We make no specific assumption on the complex coefficients $N_{\alpha
  i}$. In particular, we do not relate them to the unitary matrix
$W$, we allow them to be arbitrary (sufficiently smooth) functions of energy, and they can
be different for neutrino production and detection. But we do assume
that they are the same for different experiments (at the same energy).

Below we focus on the experimentally relevant $\nu_\mu\to\nu_\mu$
disappearance and $\nu_\mu\to\nu_e$ appearance channels. Under the
assumptions $(i),(ii),(iii)$ the corresponding probabilities are
obtained as~[PRD]
\begin{align}
  P_{\mu\alpha} =& \left | \sum_{i=1}^3 c^\alpha_i e^{-i\lambda_iL} \right|^2 
  \\
  =& \sum_i |c^\alpha_i|^2 + 2\sum_{j<i} \Re(c_i^\alpha c_j^{\alpha *})\cos(\omega_{ij}L)
  \nonumber\\
  &-2\sum_{j<i} \Im(c_i^\alpha c_j^{\alpha *})\sin(\omega_{ij}L) \,, \label{eq:prob-gen}
\end{align}
with $c_i^\alpha \equiv (N_{\alpha i}^{\rm det})^* N^{\rm prod}_{\mu i}$ and $\omega_{ij} \equiv \lambda_j-\lambda_i$. 
Similar expressions are obtained in the context of non-unitary mixing, e.g., \cite{Antusch:2006vwa,Escrihuela:2015wra,Fong:2017gke}.
As usual we
have traded the time dependence in the evolution equation with the
space coordinate, leading to the appearance of the baseline $L$ in the
above probabilities. Therefore, T reversal is formally equivalent to
$L \to -L$~\cite{Cabibbo:1977nk, Akhmedov:2001kd}. 
The first line of Eq.~\eqref{eq:prob-gen} is invariant under T, whereas the second line is T-odd. 
Fundamental TV can be established by proving the presence of the $L$-odd
term in the probability. 

For practical reasons we will introduce one more assumption. We list
it here for completeness and provide further discussion and motivation
below:

$(iv)$ We impose that the oscillation frequencies
$\omega_{ij}$ deviate only weakly from the ones 
corresponding to the standard three-flavour oscillation case.

Note that our assumptions $(i),(ii),(iii)$ are rather general and cover a large class of new physics scenarios, including non-standard interactions in production, detection, and propagation \cite{Falkowski:2019kfn}, generic
non-unitarity \cite{Antusch:2006vwa,Escrihuela:2015wra,Fong:2017gke}, or the presence of sterile neutrinos, as
long as their associated oscillation frequencies are large compared to $\omega_{ij}$ \cite{Blennow:2016jkn}. 

\bigskip\textbf{The TV test.}
The strategy we propose to probe the $L$-odd terms is to measure the
oscillation probability as a function of $L$ at a fixed energy and check whether $L$-even terms are enough to describe the data or if TV
is required.  Under these conditions, the effective frequencies and
mixings in the Hamiltonian are the same, and so the data at different
baselines (but at the same energy) can be consistently combined. 
Notice that antineutrino data cannot be analyzed together
with neutrino data, as their effective frequencies and mixings are in
general different from the neutrino's; nevertheless, separate tests could
be made for neutrino and antineutrino data.

In the absence of TV,
all $c^\alpha_i$ are real and
the data points could be described by the $L$-even part of the oscillation probability. We define ($c^\alpha_i$ real)
\begin{equation}
	P_{\mu\alpha}^\mathrm{even} (L, E; \theta) =
	\sum_{i} (c_i^{\alpha})^2 
	+2 \sum_{j<i} c_i^{\alpha} c_j^{\alpha} \cos(\omega_{ij} L) \,.
	\label{eq:ProbEven}
\end{equation}
For the two relevant channels, these probabilities depend on 8
parameters, which we collectively denote by $\theta$: 6 real
coefficients $c_i^{\mu},c_i^e$ $(i=1,2,3)$ and two independent
$\omega_{ij}$, e.g., $\omega_{21}$ and $\omega_{31}$. We assume now
that the probabilities $P_{\mu\mu}$ and $P_{\mu e}$ are measured at a
fixed energy at several baselines $L_b$. We denote the corresponding
measured values by $p_b^\mathrm{dis}$ and $p_b^\mathrm{app}$ with the
uncertainties $\sigma_b^\mathrm{dis}$ and $\sigma_b^\mathrm{app}$, respectively. Below we
are going to assume that $p_b^\mathrm{dis}$ and $p_b^\mathrm{app}$
correspond to the values predicted 
by standard three-flavour neutrino (3$\nu$) oscillations in matter.

We now ask the question if we can exclude the hypothesis of T conservation
parametrised by Eq.~\eqref{eq:ProbEven}, if the data correspond to 
$3\nu$ oscillations \emph{with} TV, i.e., for a CP phase $\delta$ different from
0 or $\pi$. To this aim we construct the
$\chi^2$ function
\begin{align}
	\nonumber
	\chi^2_\mathrm{even}(E; \theta) =& 
	\sum_{b=1}^{N_L} \left[ \frac{P^\mathrm{even}_{\mu\mu}(L_b,E; \theta) - p_b^\mathrm{dis}}{\sigma_b^\mathrm{dis}} \right]^2  \\
	&+\sum_{b=1}^{N_L} \left[ \frac{P^\mathrm{even}_{\mu e}(L_b,E; \theta) - p_b^\mathrm{app}}{\sigma_b^\mathrm{app}} \right]^2 .
	\label{eq:chi2coef}
\end{align}
The best-fit T-conserving model is obtained by considering $\chi^2_\text{min}(E)
= \text{min}_\theta\left[ \chi^2_\mathrm{even}(E;\theta)\right]$.  
We will take the value of $\chi^2_\text{min}(E)$ as a rough indication of how strongly T conservation can be excluded by data, and leave a more detailed statistical analysis for future work.
Considering that each baseline provides 2 data points (appearance and disappearance) and that the T-even model has 8 parameters, it is clear that we need more than 4 experiments at different baselines. Let us note, however, that our parameterization includes so-called zero-distance effects, due to the
non-unitary mixing in Eq.~\eqref{eq:mixing}. Therefore, the
near-detector(s) of long-baseline experiments provide already two data points at $L \approx 0$ and effectively only more than 3 experiments are needed.

This requirement can even be further relaxed if we impose one
additional assumption, which can be motivated by the fact that we have
overwhelming evidence that the standard three-flavour scenario is
approximately correct and any new physics effect can only be
sub-leading. Therefore, we introduce assumption $(iv)$ mentioned
above: we assume that the oscillation frequency $\omega_{21}$ deviates only weakly from the one corresponding to the standard $3\nu$ case. Technically we impose this requirement by calculating the effective mass-squared difference in matter, $\Delta \tilde
m^2_{21}(E)$, assuming the standard matter effect and add the following prior to Eq.~\eqref{eq:chi2coef}:
\begin{equation}
	\chi^2_\mathrm{even}(E;\theta) \mapsto 
	\chi^2_\mathrm{even}(E;\theta) +
	\left[ \frac{ \Delta \tilde m^2_{21}(E) - 2E\omega_{21}}{\sigma_{21}} \right]^2 \,.
	\label{eq:chi2prior}
\end{equation}
The frequencies $\omega_{ij}$ are determined by the effective evolution
Hamiltonian, assumption $(i)$, and are independent of the mixing in
Eq.~\eqref{eq:mixing}. A general parameterization of the Hamiltonian
is provided by the non-standard neutrino interaction scenario, see {\it e.g.,} Ref.~\cite{Esteban:2018ppq}. Using the results of Ref.~\cite{Esteban:2018ppq} we estimate the possible deviation to $\sigma_{21} = 0.1 \Delta \tilde m^2_{21}$, see the appendix for more details.
It turns out that the other independent frequency,
$\omega_{31}$, is effectively constrained by the long-baseline data
used in our fit, and therefore it is not necessary to impose an
analogous prior for it. The prior in Eq.~\eqref{eq:chi2prior} acts as an additional data point for each energy bin (note that also the prior is energy dependent). Therefore, under this additional assumption, we come to the remarkable result that our model-independent test can be performed already with 3 experiments at different baselines plus near detectors. 

The crucial requirement, however, is sufficient overlap in neutrino energy. If experiments have overlapping energy ranges, we can combine information from different energies. However, to be completely model-independent, the minimization has to be done individually for each energy, since we do not want to make any assumptions about the energy dependence of the unknown new physics. This is an important difference to usual model-dependent analyses.

\bigskip\textbf{Realistic baselines and energies.}
Let us now consider planned long-baseline accelerator experiments in
order to see if such a test realistically can be carried out in the
future. We consider the following experiments: the DUNE project in USA
($L=1300$~km) \cite{Abi:2020wmh,Abi:2020evt}, T2HK in Japan
($L=295$~km) \cite{Abe:2018uyc}, with the option of a second detector
in Korea, T2HKK ($L=1100$~km, $1.5^\circ$ off axis) \cite{Abe:2016ero}, and a long-baseline
experiment at the European Spalation Source in Sweden, ESS$\nu$SB
($L=540$~km)~\cite{Baussan:2013zcy,Blennow:2019bvl}.

\begin{figure}[t!]
	\centering
	\includegraphics[width=\columnwidth]{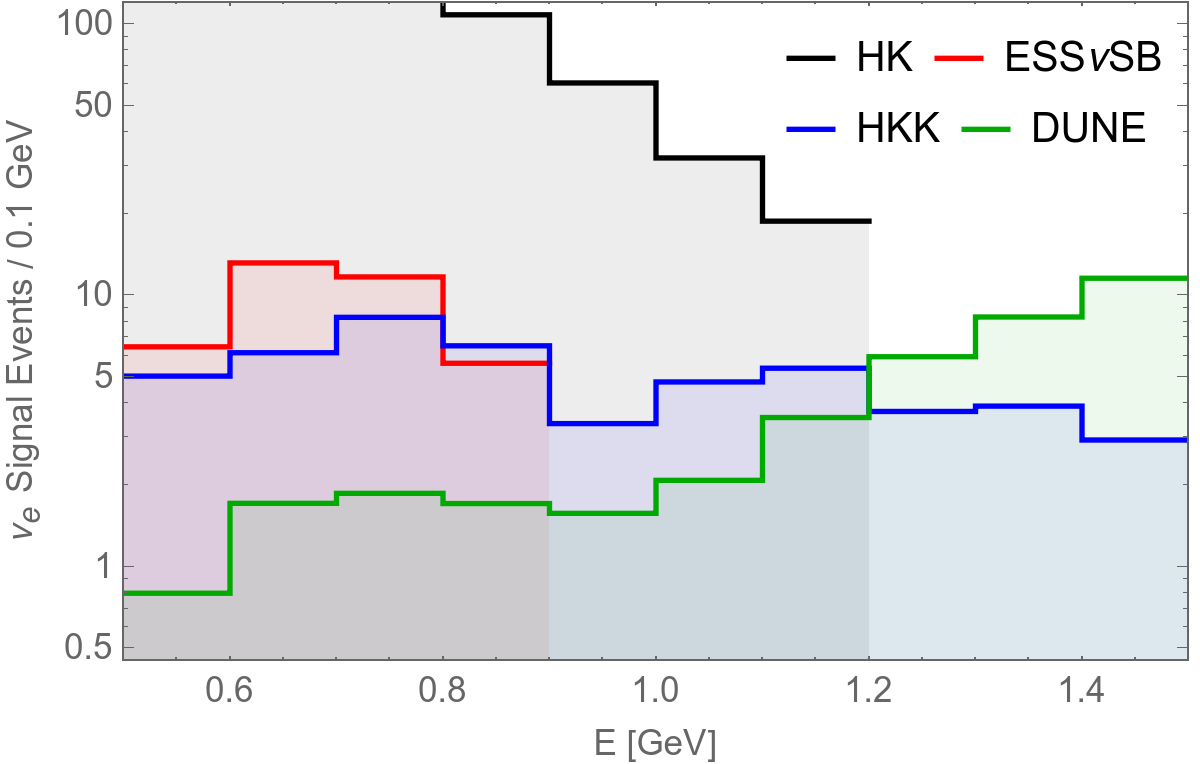}
	\caption{
	    Estimated number of appearance signal events at future accelerator experiments,
        assuming normal mass ordering and true $\delta = 90^\circ$. 
        Data from Refs.~\cite{Acciarri:2015uup, DeRomeri:2016qwo}~(DUNE), 
        \cite{Abe:2018uyc}~(T2HK),
        \cite{Abe:2016ero}~(T2HKK),
        and \cite{Blennow:2019bvl}~(ESS$\nu$SB).
        }
	\label{fig:expData}
\end{figure}

Expected event numbers are obtained from Design Reports or detailed studies of the physics potential and are shown for the appearance channel in the case of  $3\nu$ oscillations and $\delta = 90^\circ$ in Fig.~\ref{fig:expData}. In practice, we will see that only the two energy bins between 0.7 and 0.9~GeV provide relevant sensitivity, as data points with sufficient statistics are needed at 1st and 2nd oscillation maxima. We note that the
energy spectrum from the NO$\nu$A experiment \cite{Acero:2019ksn} has
no overlap with the T2K beam and therefore it cannot be used for this
analysis. We use the information from Fig.~\ref{fig:expData} (and the corresponding data for the disappearance channel) to estimate
the statistical uncertainties in Eq.~\eqref{eq:chi2coef} as
$\sigma_{br}/P^{\rm even}(L_b, E_r) = \sqrt{S_{br}+B_{br}}/S_{br}$ at baseline $b$ and energy bin $r$.
We take the background events $B_{br}$ directly from the experimental studies
and estimate the number of signal events from the $N_{br}$ in the Figure assuming $S_{br} = N_{br} \times P^{\rm even}(L_b, E_r; \theta) / P^{3\nu}(L_b, E_r)$.
For the near detector data points,
we assume the standard $P_{\alpha\beta}(L\to 0)=\delta_{\alpha\beta}$ with $\sigma = 0.01$.

\begin{figure}[t]
	\centering
	\begin{tikzpicture}
	    \def\x{0.208\textwidth}
	    \def\y{-3.13}
	    \def\w{0.25\textwidth}
	    \node at (\x,0){\includegraphics[width=\w]{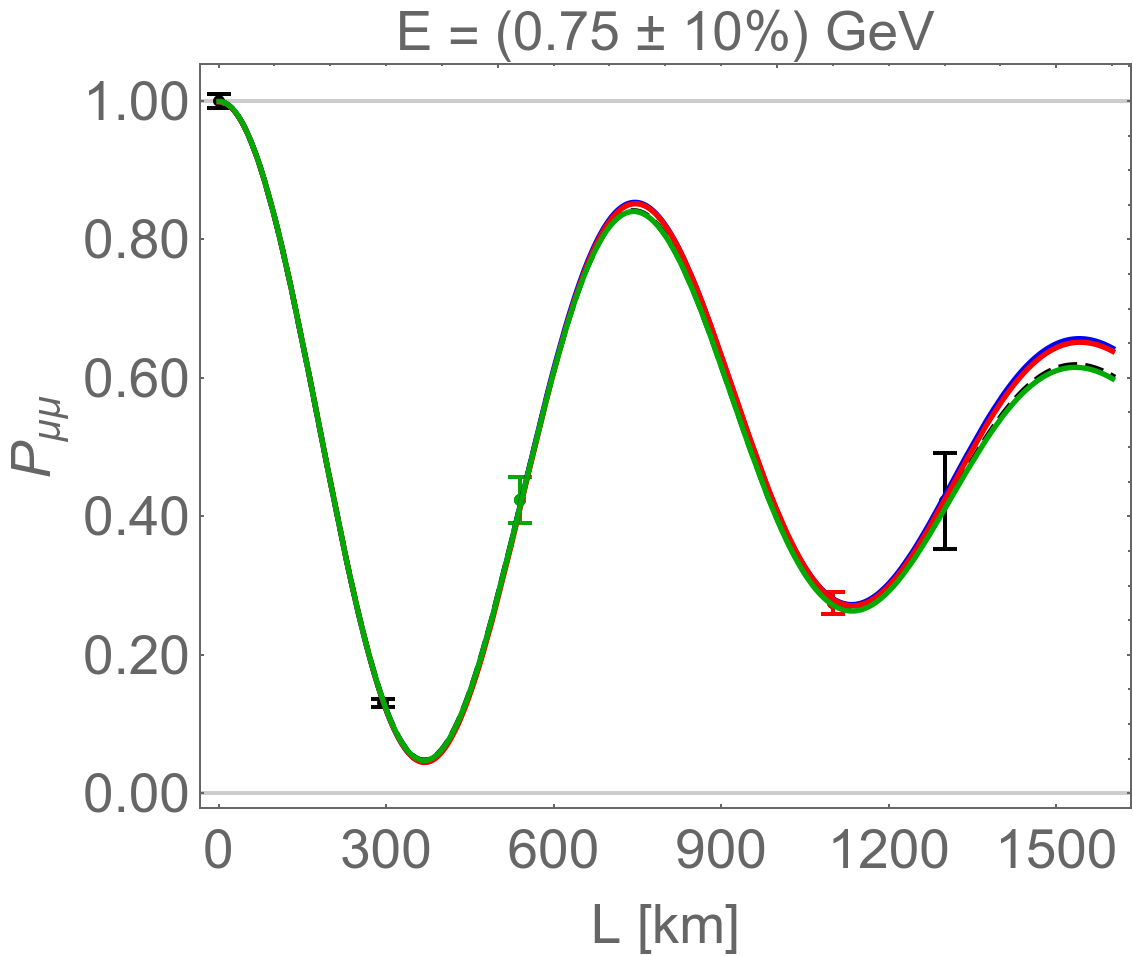}};
	    \node at (0,0){\includegraphics[width=\w]{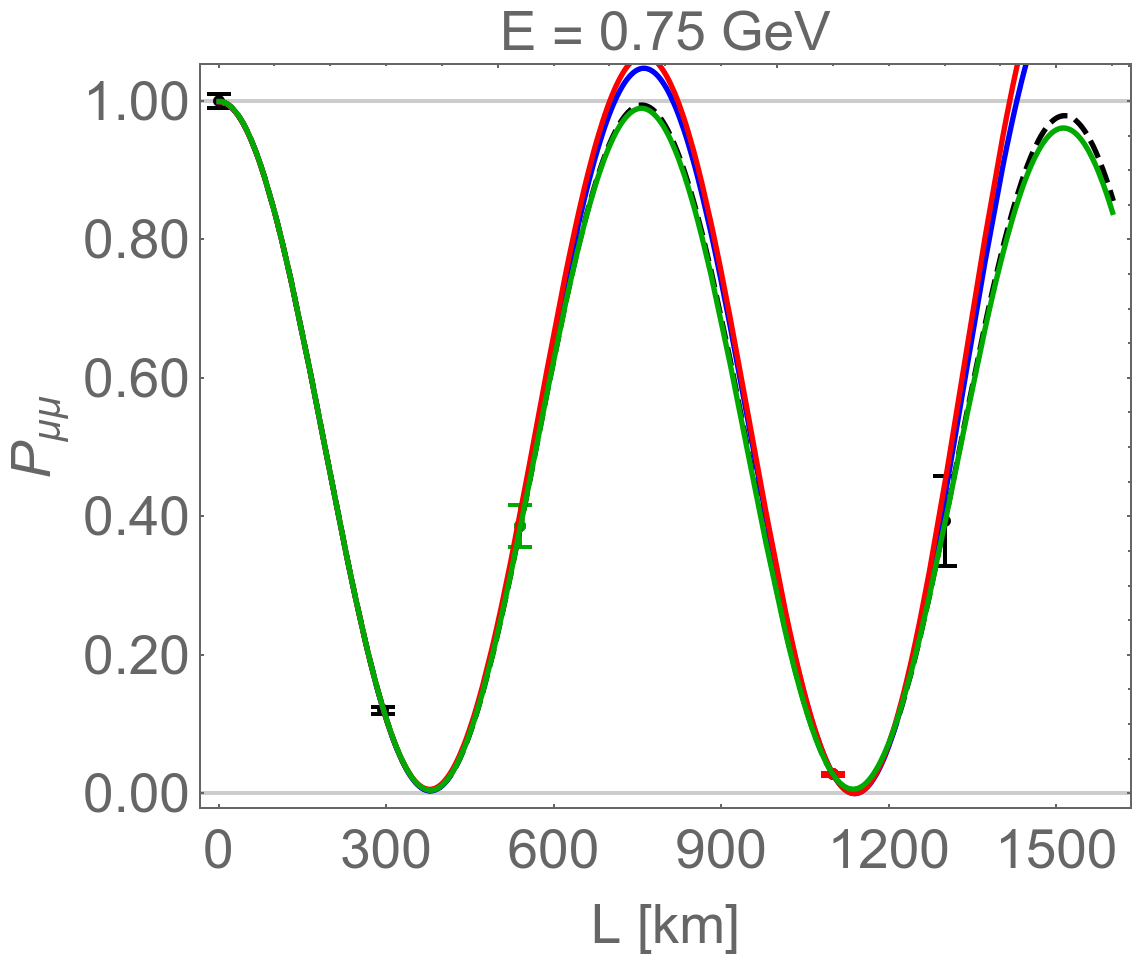}};
	    
	    \begin{scope}[shift={(0,\y)}]
	        \node at (\x,0){\includegraphics[width=\w]{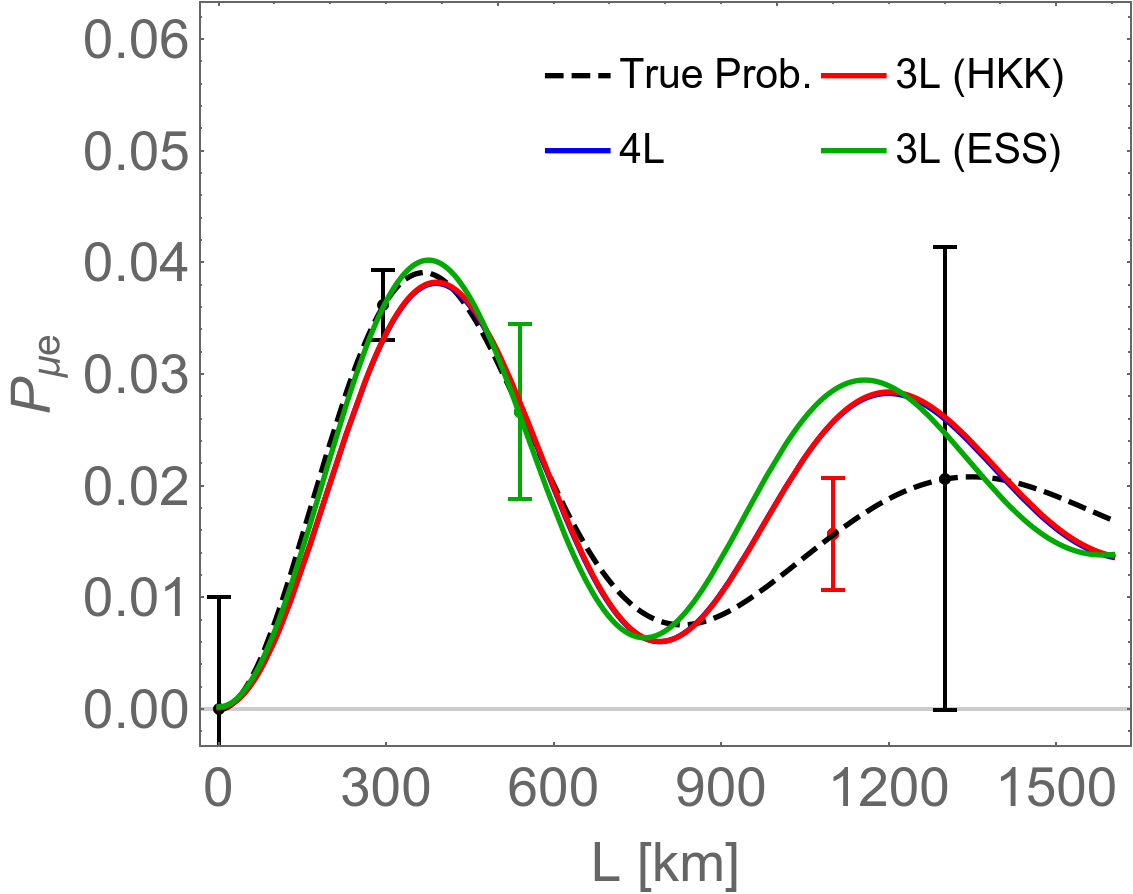}};
	        \node at (0,0){\includegraphics[width=\w]{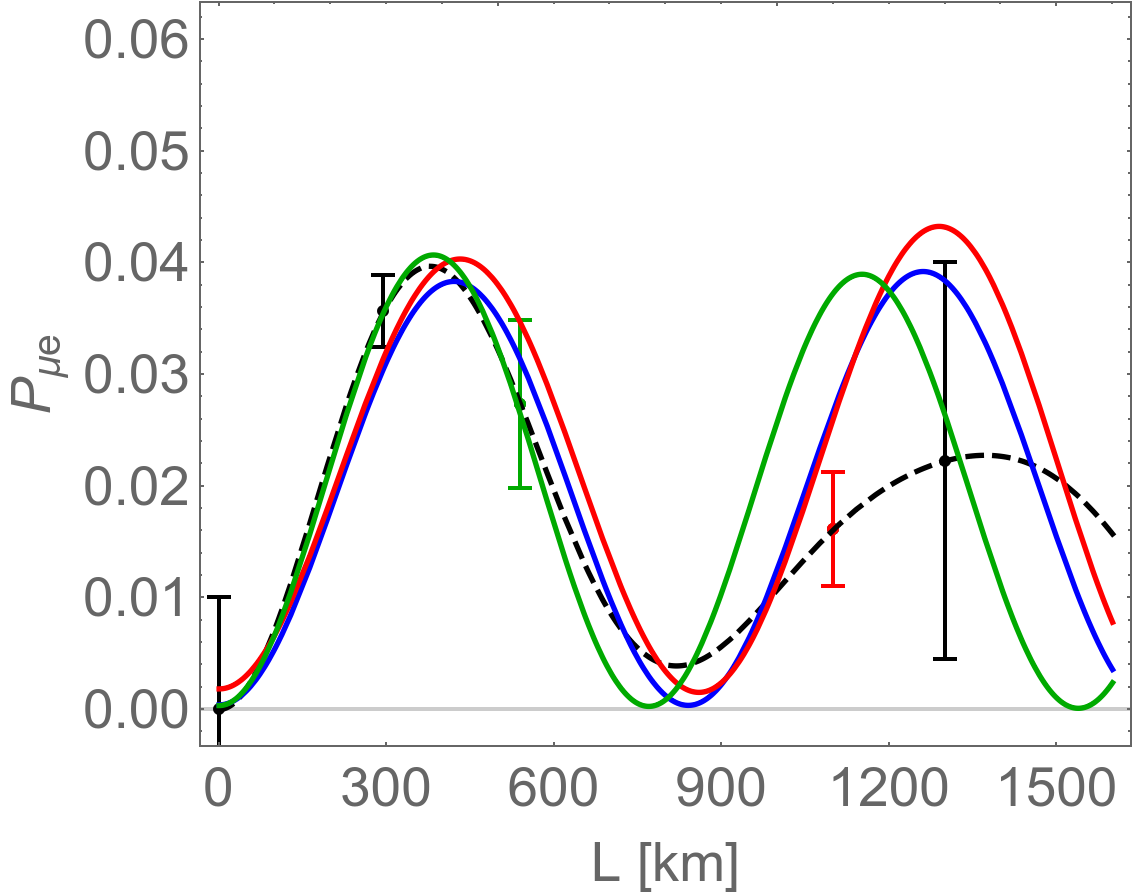}};
	    \end{scope}
	\end{tikzpicture}

	\caption{
	    Data points for the disappearance (top) and
        appearance (bottom) channels at the baselines of DUNE, T2HK,
        T2HKK, ESS$\nu$SB and a near detector location for $E=0.75$~GeV.
        Data points are generated for standard three-flavour oscillations in matter with
        normal mass ordering and $\delta = 90^\circ$, and the corresponding oscillation
        probability is shown as black-dashed. Error bars show $1\sigma$
        statistical errors. The solid curves show
        the best-fit model-independent $L$-even probabilities using all baselines (4L, blue),
        DUNE + T2HK + T2HKK (3L (HKK), red),
        or DUNE + T2HK + ESS$\nu$SB (3L (ESS), green). Left (right) panels are without (with)
        the smearing due to a $10\%$ energy resolution.    
    }
    \label{fig:bestfitProb}
\end{figure}

In Fig.~\ref{fig:bestfitProb} we show the data points for the
appearance and disappearance probabilities as a function of the
baseline for the 0.7--0.8 GeV energy bin. We can see that the disappearance
data points essentially fix the oscillation frequency, whereas the
appearance data are crucial for the TV test. The ``true'' oscillation
probability assumed to generate the data points correspond to standard
$3\nu$ oscillations with maximal TV ($\delta = 90^\circ$) 
and normal mass ordering.
We find that no satisfactory $L$-even fit is possible for the 4L and 3L~(HKK) combinations at this energy.
The essential information is obtained from the relative heights of the first and second appearance oscillation peaks, see the appendix for further discussion. 
Note that disappearance probabilities can reach values larger than one in our fit, since we do not impose unitarity in our effective parameterization of the T-even transitions.
 
In order to connect our test with experiments,
one should take into account the fact that finite energy resolution
effectively changes the $L$-dependence in their measurements,
which will in turn affect the sensitivity of the TV test.
We assume a given energy resolution $\Delta E$ around the central bin energy $E_0$, and smear the transition probability by convoluting it with
a Gaussian with mean $E_0$ and width $\Delta E$. To illustrate the effect we assume here $\Delta E = 0.1 E_0$. In order to perform the convolution one must assume a certain energy dependence of the transition probability. Our assumption is that the energy dependence of the amplitudes $c_i^\alpha$ is slow enough, such that it can be neglected within an interval of few $\Delta E$. The only significant energy dependence would thus be in the oscillation phases $\omega_{ij}$. According to assumption $(iv)$ introduced above, we assume that $\omega_{31} \propto 1/E$, as in the standard $3\nu$ oscillation case. We have checked that our results are independent of energy smearing of $\omega_{21}$ terms. The impact of the finite energy resolution is illustrated in the right panels of Figure~\ref{fig:bestfitProb}.

\begin{table}[t!]
  \caption{
    Fit to data with the $\Delta m^2_{21}$ prior $\sigma_{21} = 0.1$ in Eq.~(\ref{eq:chi2prior})
    assuming normal mass ordering 
    and a true $\delta = 90^\circ$.
	Units of $E$ are GeV. Columns correspond to different combinations of DUNE, T2HK, T2HKK, ESS$\nu$SB.
	The values outside (inside) the brackets show the $\min(\chi^2)$
	without (with) smearing the data with a $10\%$ energy resolution.
  }
  \label{tab:chi2bins}
  \setlength{\tabcolsep}{3pt}
  \begin{tabular}{ccccc}
  \hline\hline
    $E$   &w/o HKK       &w/o DUNE      &w/o ESS        &all  \\
  \hline
    0.65        &0.07 [0.03] &0.76 [0.65] &0.04 [0.21]  &0.79 [0.67]    \\
    0.75        &0.04 [0.04] &6.95 [4.78] &7.92 [4.82]  &8.60 [4.86]    \\
    0.85        &0.54 [0.53] &0.76 [2.18] &2.75 [2.96]  &3.15 [3.06]    \\
    0.95        &-           &-           &0.42 [0.98]  &-    \\
  \hline
    Tot.       &0.65 [0.60] &8.46 [7.60] &11.13 [8.97] &12.54 [8.59]   \\
  \hline\hline
  \end{tabular}
\end{table}

Our results for maximal TV are summarized in Tab.~\ref{tab:chi2bins}, which shows the $\chi^2_{\rm min}$ values for the various energy bins for different experiment combinations, with and without including the energy smearing. We observe that 0.75~GeV is the most relevant energy bin, whereas the one at 0.85~GeV still provides some sensitivity. The strong impact of the energy resolution is apparent. We also find that the detector in Korea is essential, 
whereas both DUNE and ESS provide little sensitivity but at least one of them is needed to fix the $\omega_{ij}$ from disappearance data.

\begin{figure}[t]
	\centering
	\begin{tikzpicture}
	    \def\h{5.5cm}
	    \node at (0,0){\includegraphics[height=\h]{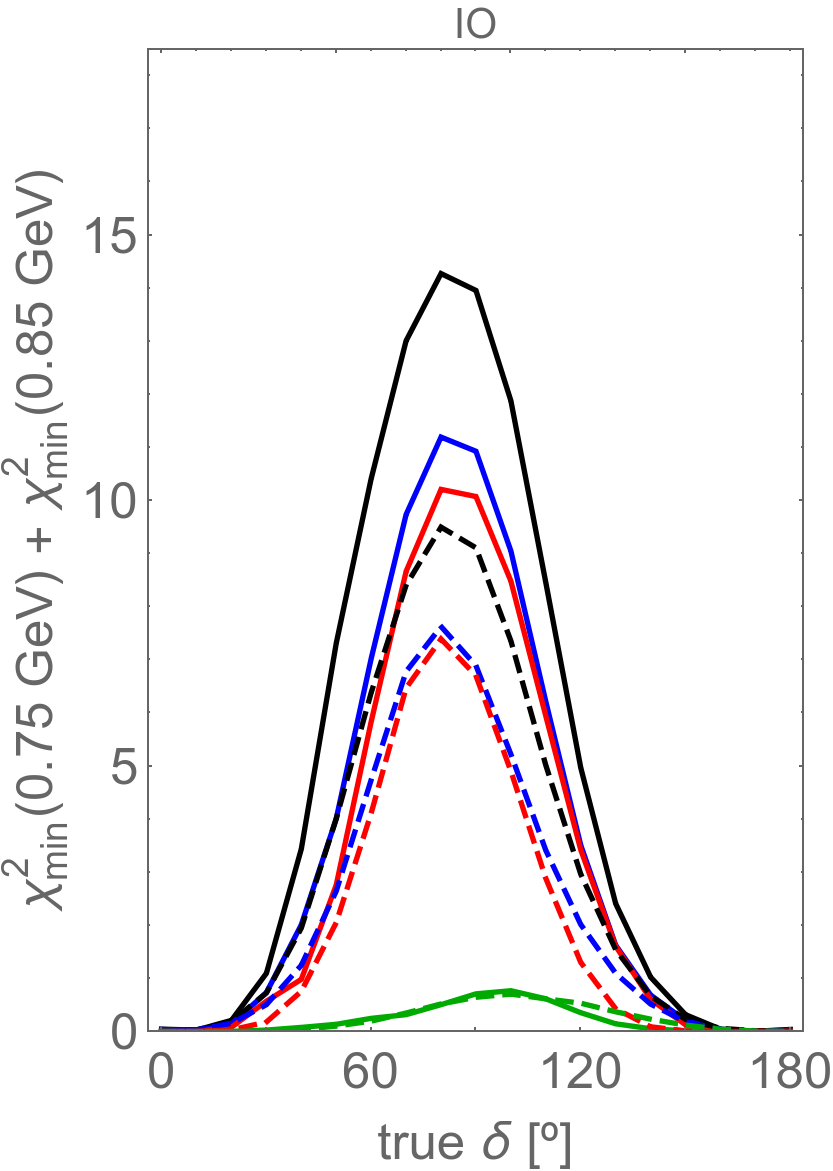}};
	    \node at (-4,0){\includegraphics[height=\h]{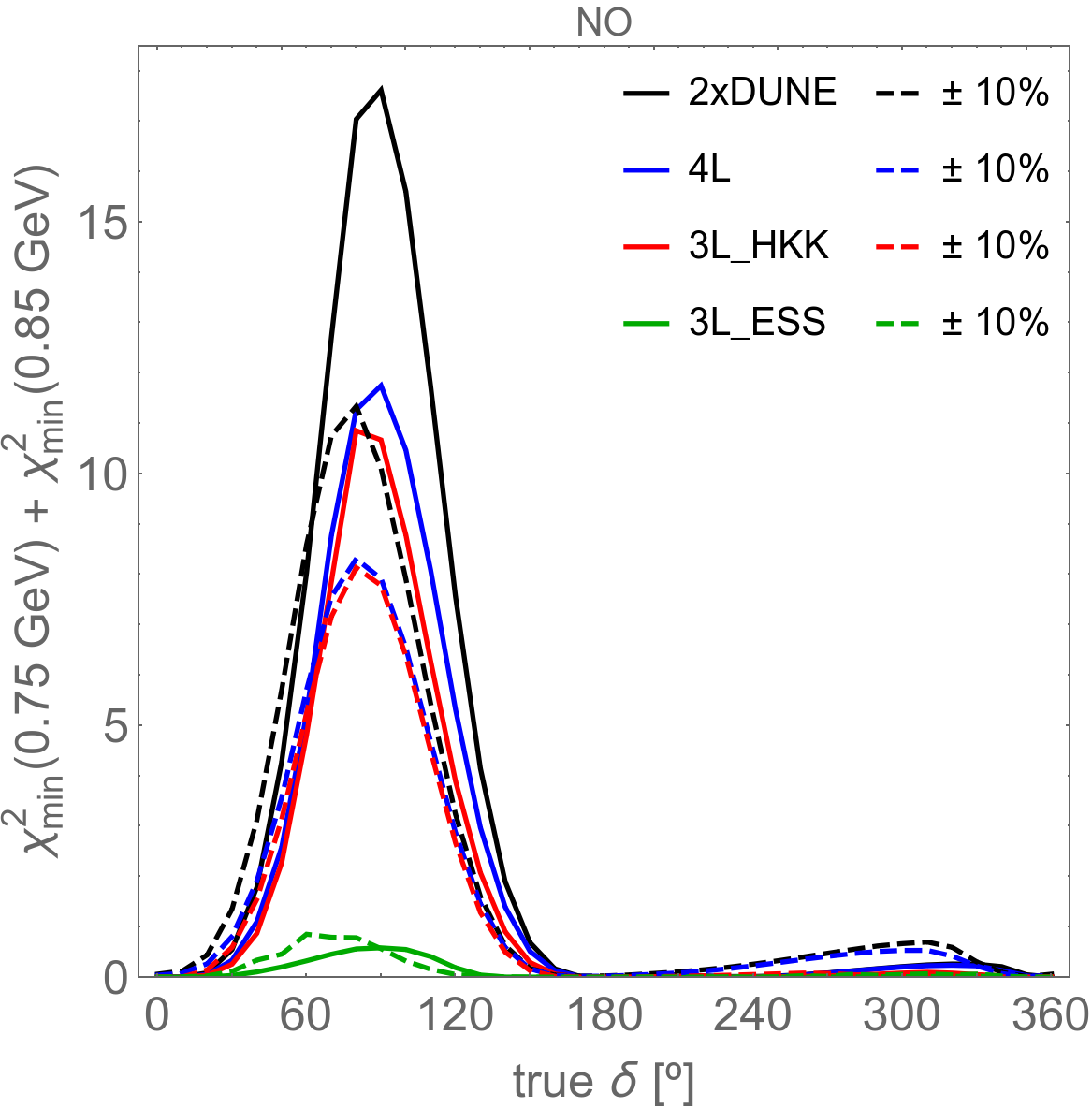}};
	\end{tikzpicture}
	\caption{$\chi^2_{\rm min}$ summed for the energy bins around 0.75 and 0.85~GeV, 
	    with perfect (solid) or $10\%$ (dashed) energy resolution.
	    We show the fit to all 4 experimental baselines (4L),
	    DUNE + T2HK + T2HKK (3L (HKK)),
	    and DUNE + T2HK + ESS$\nu$SB (3L (ESS)),
	    as well as the effect in 4L of DUNE having twice as many events (2xDUNE).
	    Neutrino data is assumed, with
	    normal (inverted) mass ordering for the left (right) panel.
	}
	\label{fig:delta}
\end{figure}

In Fig.~\ref{fig:delta} we show the summed $\chi^2_{\rm min}$ contributions from the 0.75 and 0.85~GeV bins as a function of the value of the $3\nu$ CP phase $\delta$ assumed to calculate the ``data'' to which the T-even model is fitted. 
In addition to the features mentioned above, we see from Fig.~\ref{fig:delta} that the test is sensitive only to $\delta\simeq 90^\circ$, whereas no sensitivity appears around $270^\circ$. 
This behaviour stems from the enhancement of the second oscillation maximum in the latter case (contrary to its suppression around $90^\circ$):
only when the second oscillation maximum is smaller than the first one 
does the $P_{\mu e}^\mathrm{even}(L)$ fail to fit the data.
Bins with $E>1$~GeV are not useful in the test because of the absence of measurements at both maxima. See the appendix for further discussion.
For illustration purpose we show in Fig.~\ref{fig:delta} the effect doubling the event numbers in DUNE. This shows that there is significant potential to increase the sensitivity of the test by suitable optimizations. The increased sensitivity emerges from the 0.85~GeV bin, since at this energy the DUNE baseline is close to the 2nd oscillation maximum.

The results for inverted mass ordering (IO) are qualitatively similar to the one from normal ordering
(for IO we show only the relevant range of $\delta$ in Fig.~\ref{fig:delta}). Further details on IO are given in the appendix.
If antineutrino data are assumed (instead of neutrino data) the result is roughly obtained for $\delta\to2\pi-\delta$ in Fig.~\ref{fig:delta}, with highest sensitivity around $\delta \simeq 270^\circ$. This is to be expected, since antineutrino oscillation probabilities are obtained from the neutrino ones by replacing $\delta\to -\delta$ (in addition to the sign-flip of the matter potential). 


\bigskip\textbf{Summary.}
We propose a largely model-independent test to search for T violation in neutrino oscillations by comparing transition probabilities at the same energy and different baselines. 
The test can be done under rather general assumptions covering a wide range of new physics scenarios. 
Within some modest assumptions, the test can be performed already with experiments at three different baselines plus near detectors. 
The crucial requirements are sufficient event numbers in the neutrino energy overlap region between the experiments and good neutrino energy reconstruction \cite{DeRomeri:2016qwo,Chatterjee:2021wac}. 
Our estimates show that with the planned long-baseline experiments DUNE, T2HK, and T2HKK, this test can be potentially carried out. 
In order to cover all T-violating values of $\delta$, data for neutrinos and antineutrinos are necessary.
We stress that a detector at the Tokai-Korea baseline is required in addition to DUNE and T2HK.
Some optimization studies, especially in the low-energy region of the DUNE and high-energy region of the T2HKK beams, may be required. 
The results presented here warrant more detailed sensitivity studies based on realistic experiment simulations and statistical analyses, which we leave for future work.

\begin{acknowledgments}
This project has received support from the European Union’s Horizon
2020 research and innovation programme under the Marie
Sklodowska-Curie grant agreement No 860881-HIDDeN,
and from the Alexander von Humboldt Foundation. 
\end{acknowledgments}

\bibliography{Bibliography}

\newpage
\section*{Supplemenatry Material}

\subsection{Estimation of the $\Delta m_{21}^2$ prior}

If the neutrino data come from a theory with T violation,
the $L$-even fit converges to oscillation frequencies different from the real ones
trying to compensate the wrong functional form.
These different oscillation frequencies thus show either the presence of T violation
or an underlying BSM mechanism that shifts the effective mass-squared differences $\Delta \tilde m_{ij}^2$ from their (energy-dependent) standard values in matter.

We characterize the possible size of this second effect 
using the constraints on non-standard neutrino interactions
from current global oscillation data~\cite{Escrihuela:2015wra}.
The NSI effects can be parametrized in the matter part of the neutrino Hamiltonian as
\begin{equation}
    H_\mathrm{mat} = \sqrt{2} G_F N_e(x)
    \mqty[1+\mathcal{E}^{}_{ee}(x) &\mathcal{E}^{}_{e\mu}(x) &\mathcal{E}^{}_{e\tau}(x)\\
        \mathcal{E}^{}_{\mu e}(x) &\mathcal{E}^{}_{\mu\mu}(x) &\mathcal{E}^{}_{\mu\tau}(x)\\
        \mathcal{E}^{}_{\tau e}(x) &\mathcal{E}^{}_{\tau\mu}(x) &\mathcal{E}^{}_{\tau\tau}(x)] \,,
\end{equation}
where $\mathcal{E}^{}_{\alpha\beta}(x) = \epsilon_{\alpha\beta}^{e} + \epsilon_{\alpha\beta}^{p} + Y_n(x) \epsilon_{\alpha\beta}^{n}$,
$Y_n$ is the neutron-to-electron number ratio,
and $\epsilon_{\alpha\beta}^{q}$ are the non-standard couplings to electron, protons, neutrons for $q=e,p,n$, respectively.

Considering a typical value $\epsilon_{\alpha\beta}^{q} \lesssim 0.1$ from the global fit results in Ref.~\cite{Escrihuela:2015wra},
we generate random values for the non-standard couplings 
and diagonalize the Hamiltonian to study their effect on the neutrino mass squared differences.
We find that $\Delta \tilde m_{21}^2(E)$ is shifted around its standard value in matter with $\sigma = 0.09$.
Therefore, we conservatively choose $\sigma_{21}(E) = 10\% \Delta \tilde m_{21}^2(E)$ to describe 
the allowed deviations of the smallest oscillation frequency from the SM value 
via the prior term Eq.~\eqref{eq:chi2prior}. 

In the case of the large oscillation frequency,
we find that such effect is negligible:
the T-even disappearance data always fix its best-fit point to a value within the range allowed by NSIs.

Let us emphasize that this approach remains largely model independent. We consider the NSI as an effective parametrization to study how much the eigenvalues of the Hamiltonian are allowed to deviate from their standard values. This argument does not depend on the specific new physics scenario and is independent of the mixing coefficients in Eq.~\eqref{eq:mixing}.

\subsection{Discussion of particular behaviours of the $L$-even fit}

The consistency of the test requires that $\chi^2_{\rm min} = 0$ in the absence of T violation,
and we checked that we indeed recover this result whenever data are generated for $3\nu$ oscillations with $\sin\delta = 0$.
However, depending on the specific values of the data points to be fitted, it may happen that one obtains a small $\chi^2_{\rm min}$ even in the presence of T violation.
In this Section, we present some cases that illustrate the typical behaviour of the $L$-even fit.

\begin{figure*}[t!]
	\centering
	\begin{tikzpicture}
	    \def\x{0.304\textwidth}
	    \def\y{-4.6}
	    \node at (\x,0){\includegraphics[width=0.365\textwidth]{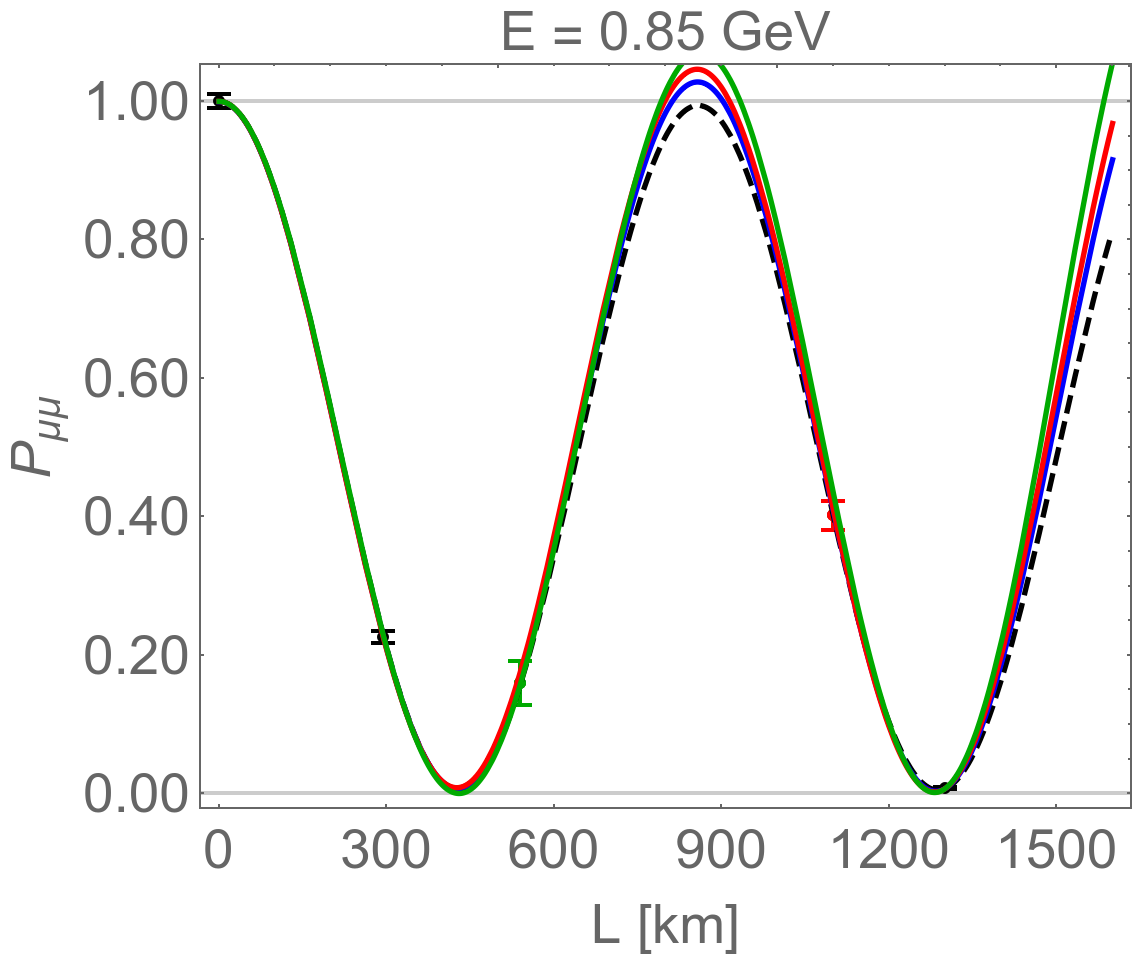}};
	    \node at (0,0){\includegraphics[width=0.365\textwidth]{img/bestfitProbs_smearing0_delta90_energy075_Dis.png}};
	    \node at (-\x,0){\includegraphics[width=0.365\textwidth]{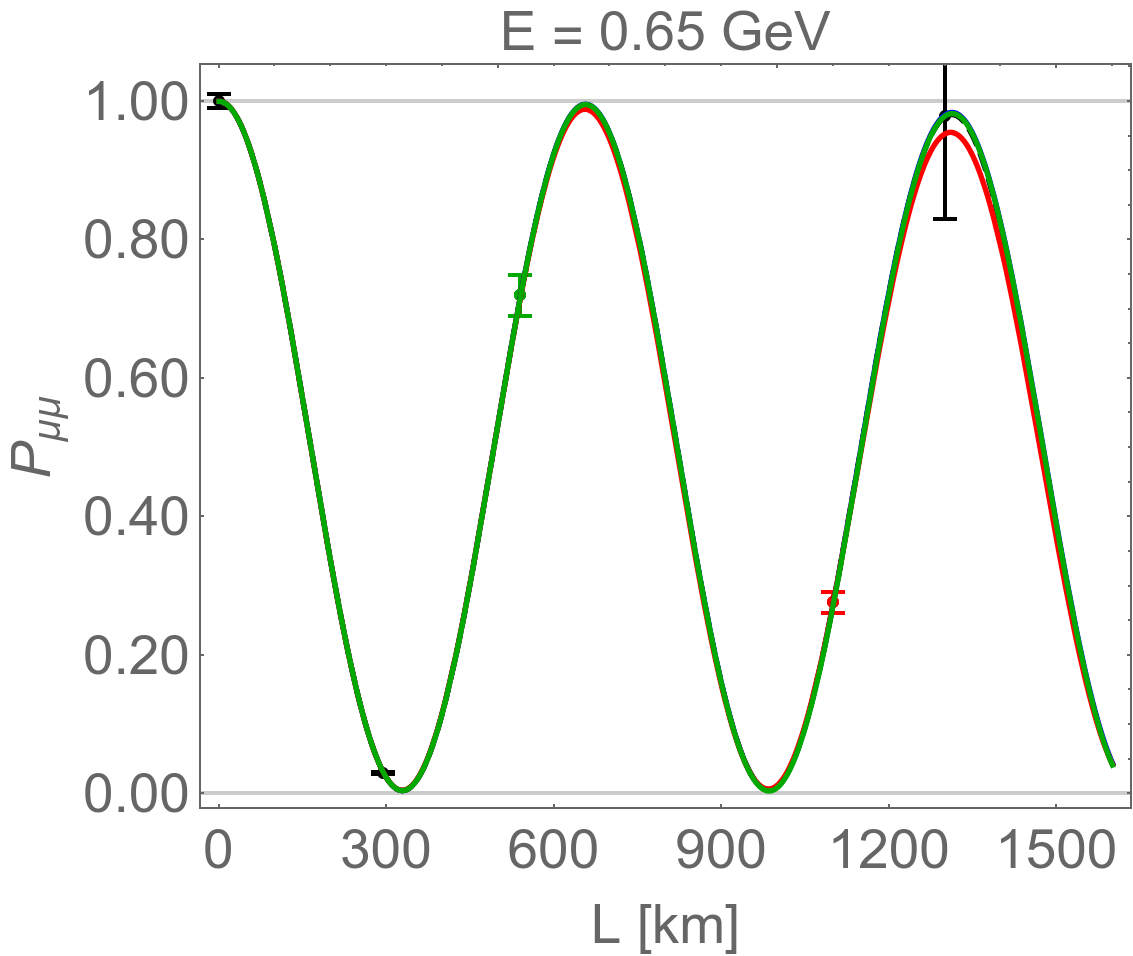}};
	    
	    \begin{scope}[shift={(0,\y)}]
	        \node at (\x,0){\includegraphics[width=0.365\textwidth]{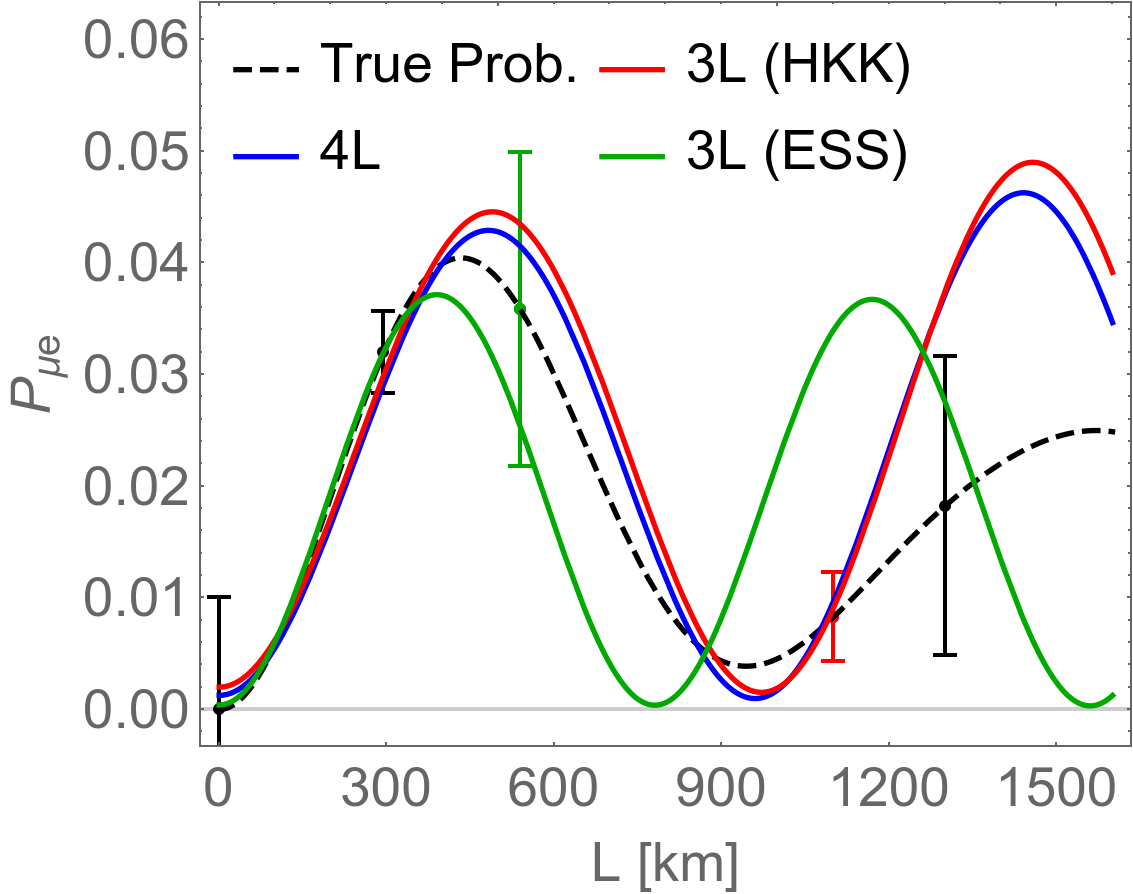}};
	        \node at (0,0){\includegraphics[width=0.365\textwidth]{img/bestfitProbs_smearing0_delta90_energy075_App.png}};
	        \node at (-\x,0){\includegraphics[width=0.365\textwidth]{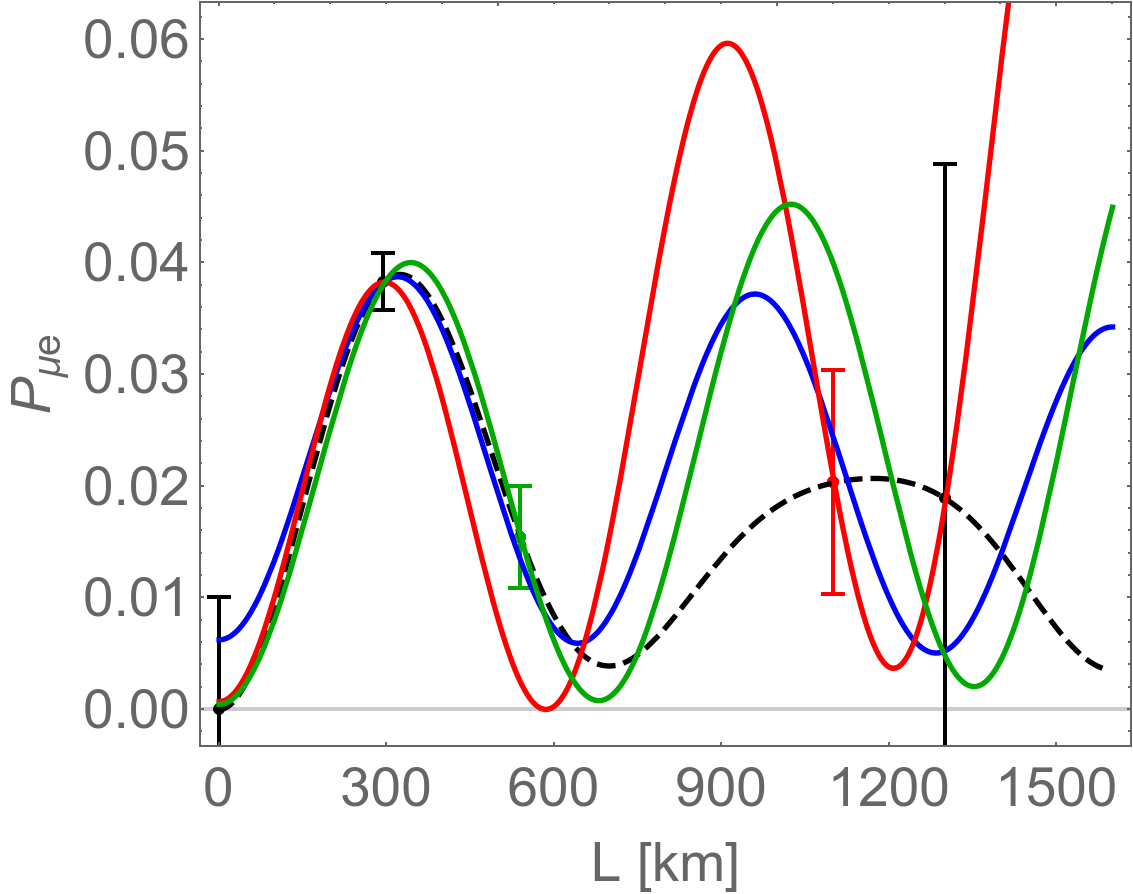}};
	    \end{scope}
	\end{tikzpicture}

	\caption{
	    Data points for the disappearance (top) and
        appearance (bottom) channels at the baselines of DUNE, T2HK,
        T2HKK, ESS$\nu$SB and a near detector location for neutrino
        energies $E=0.65$~GeV (left), 0.75~GeV (middle), and 0.85~GeV (right). 
        Data points are generated for standard three-flavour oscillations in matter with
        normal mass ordering and $\delta = 90^\circ$, and the corresponding oscillation
        probability is shown as black-dashed. Error bars show $1\sigma$
        statistical errors.
        The solid curves show
        the best-fit model-independent $L$-even probabilities using all baselines (4L, blue),
        DUNE + T2HK + T2HKK (3L (HKK), red),
        or DUNE + T2HK + ESS$\nu$SB (3L (ESS), green). We assume perfect energy resolution.
        }
        \label{fig:FitEnergies}
\end{figure*}

\begin{figure*}[t!]
	\centering
	\begin{tikzpicture}
	    \def\x{0.304\textwidth}
	    \def\y{-4.6}
	    \node at (\x,0){\includegraphics[width=0.365\textwidth]{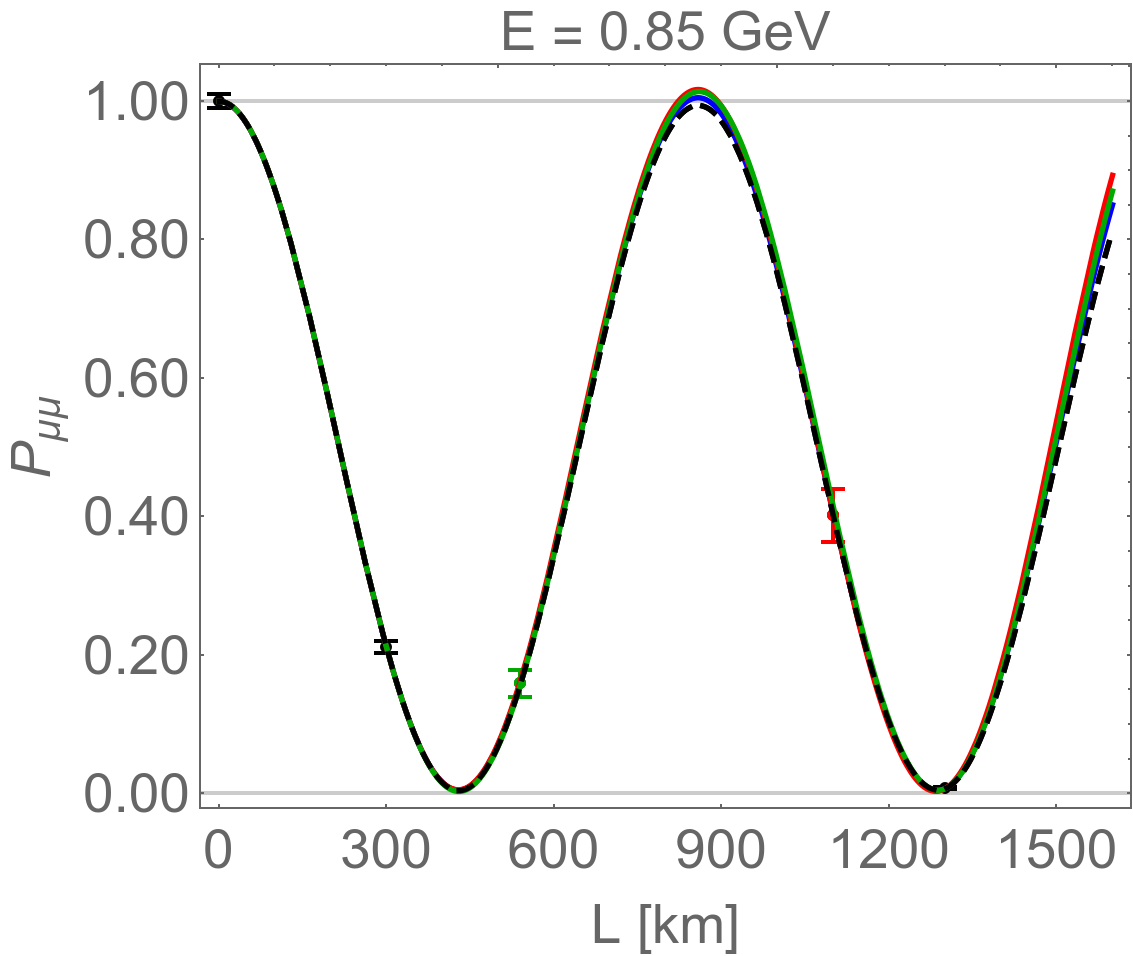}};
	    \node at (0,0){\includegraphics[width=0.365\textwidth]{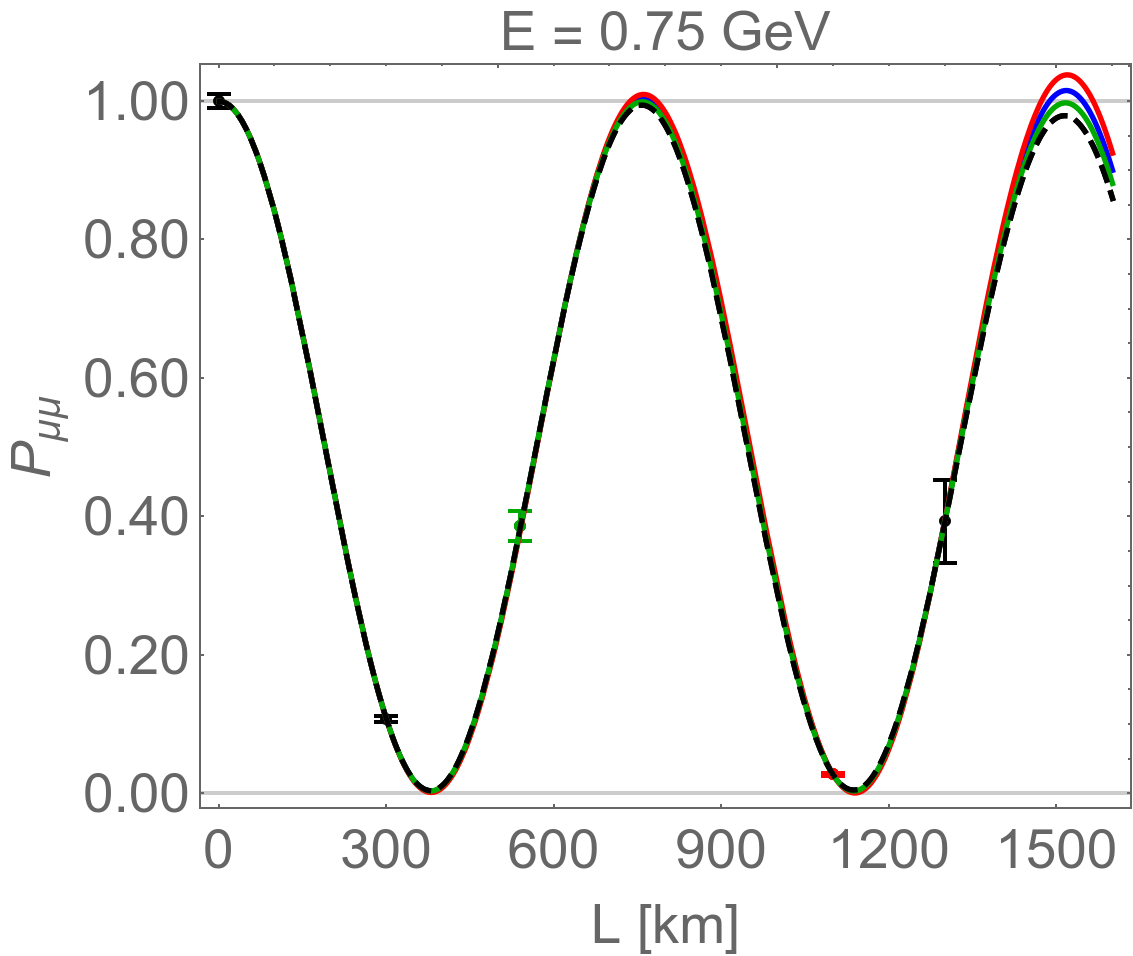}};
	    \node at (-\x,0){\includegraphics[width=0.365\textwidth]{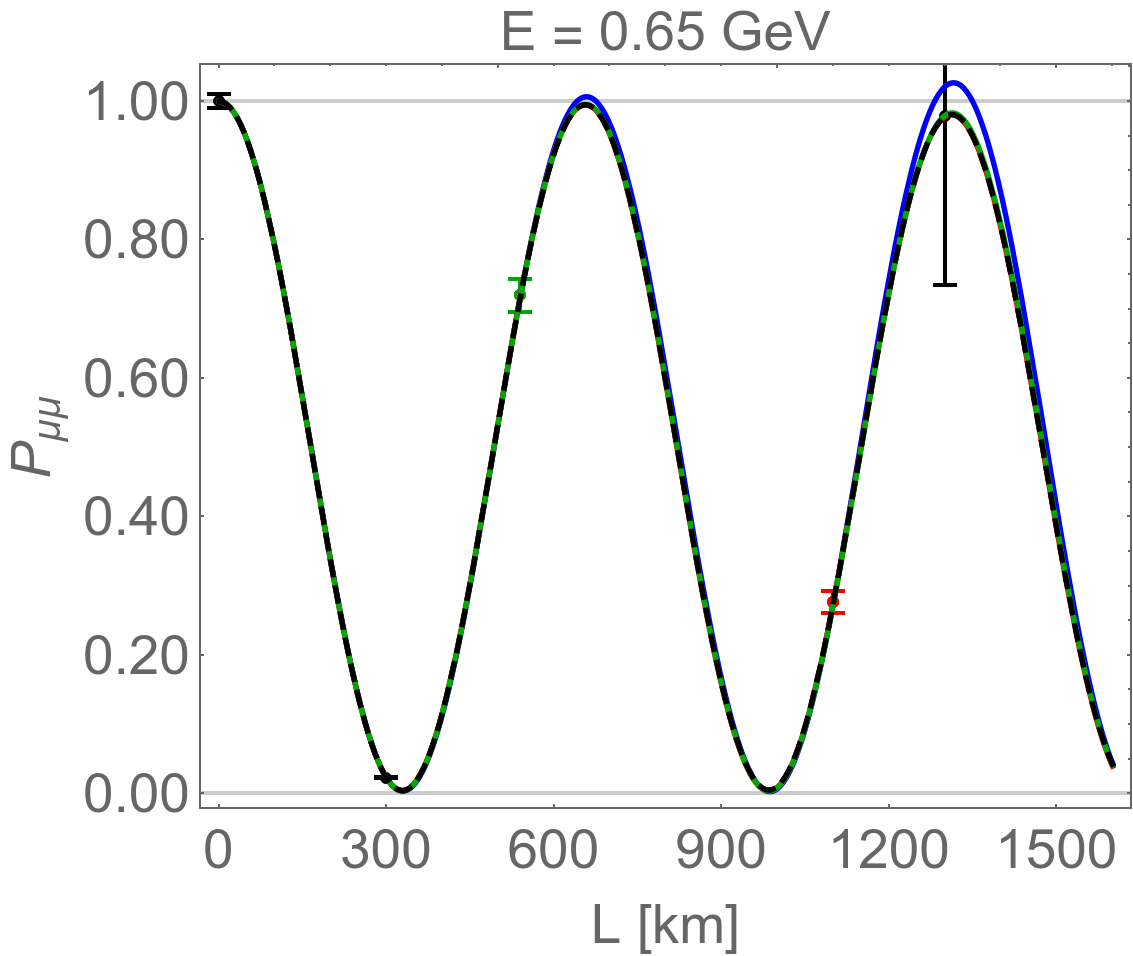}};
	    
	    \begin{scope}[shift={(0,\y)}]
	        \node at (\x,0){\includegraphics[width=0.365\textwidth]{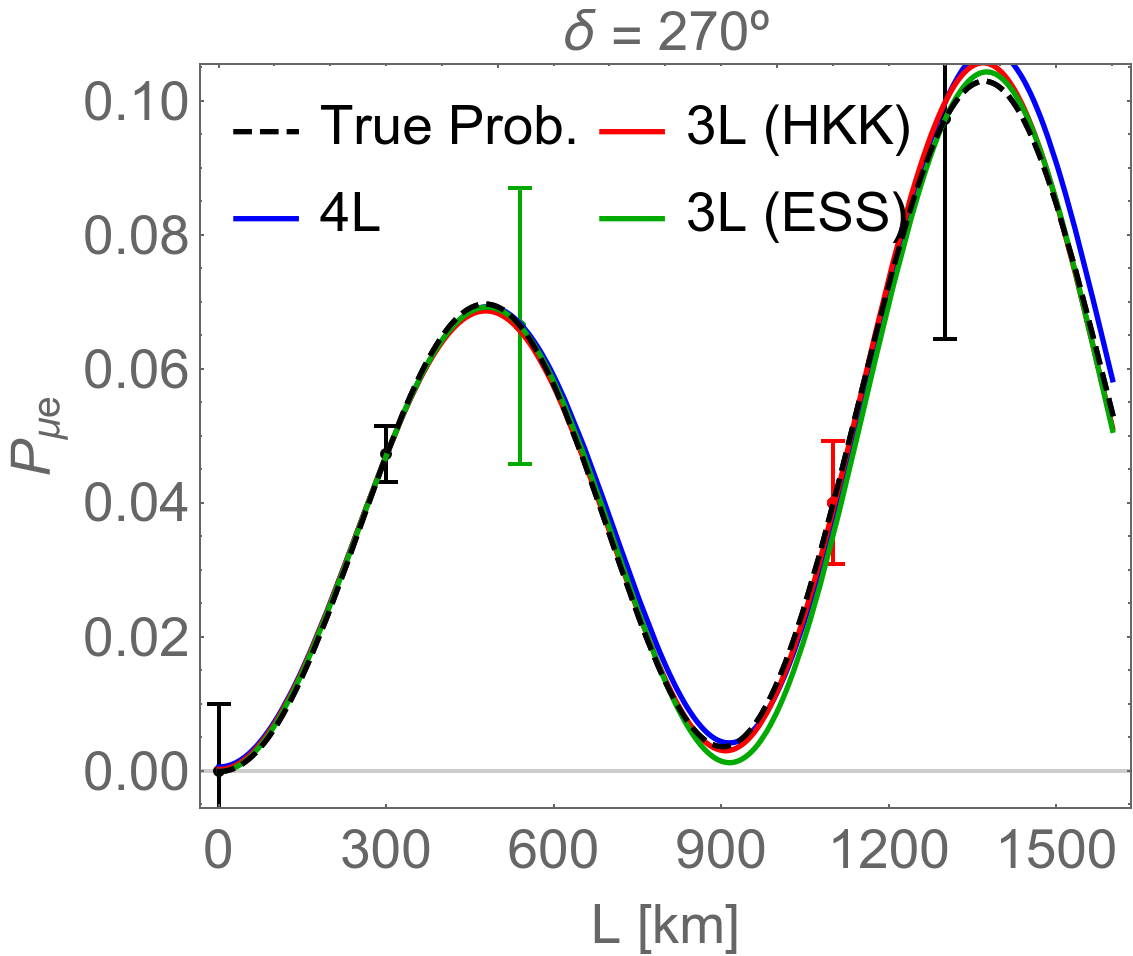}};
	        \node at (0,0){\includegraphics[width=0.365\textwidth]{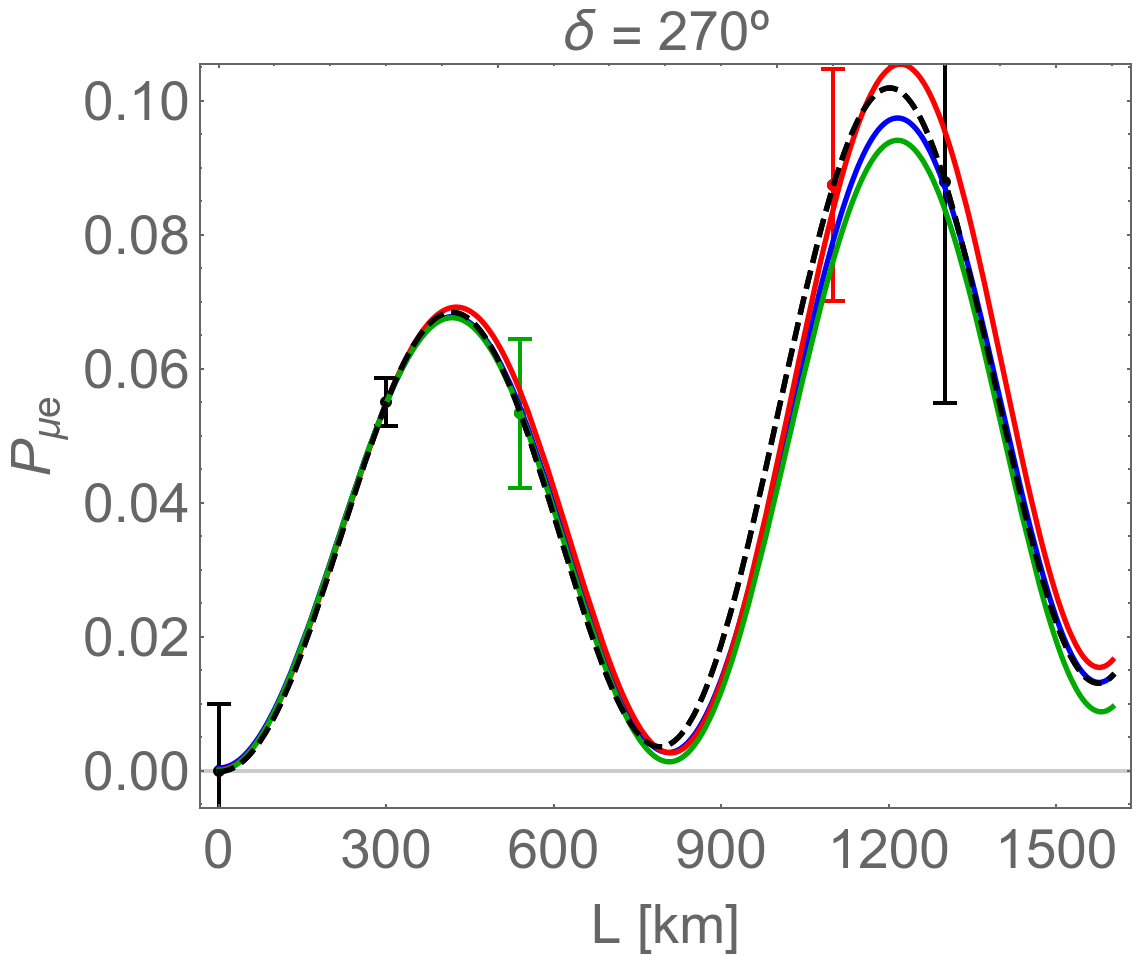}};
	        \node at (-\x,0){\includegraphics[width=0.365\textwidth]{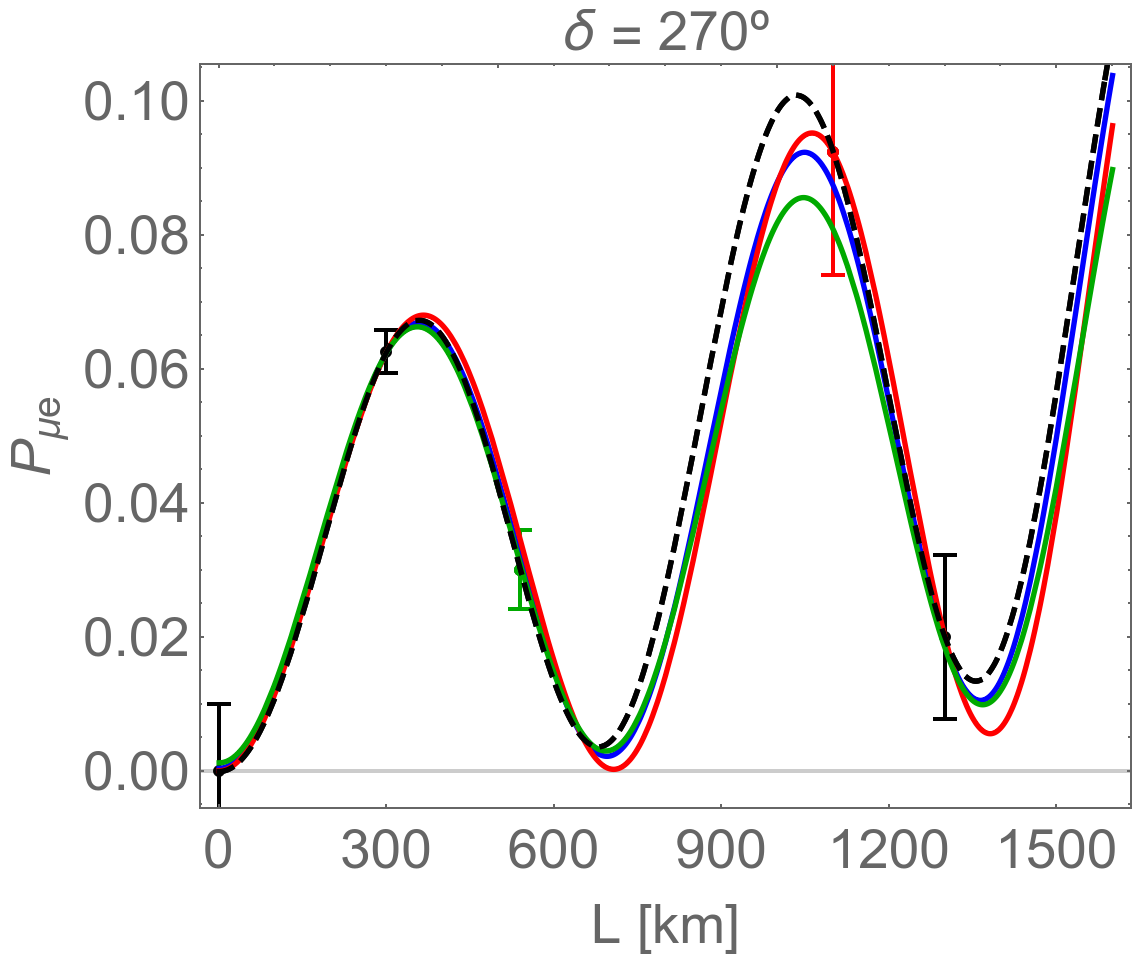}};
	    \end{scope}
	\end{tikzpicture}

	\caption{
	    Same as Fig.~\ref{fig:FitEnergies} for $\delta = 270^\circ$.
	    }
        \label{fig:FitEnergies270}
\end{figure*}

We show in Figure~\ref{fig:FitEnergies} the data and best-fit curves for $\delta = 90^\circ$ at the energies $E = 0.65, 0.75, 0.85$~GeV.
For the clarity of the following discussion,
we show the results without the smearing effect due to a finite energy resolution; the qualitative features apply also to the case including energy smearing. The middle panels of 
Figure~\ref{fig:FitEnergies} are identical to the left panels of Figure~\ref{fig:bestfitProb}. 
In general,
the disappearance data fix the largest oscillation frequency,
whereas appearance data provide the sensitivity to TV.
The two higher-energy bins are the most sensitive ones,
where the fit is unable to describe the data from the T-violating oscillation probability. 
On the other hand, we see that for $E=0.65$~GeV (left panels) the $L$-even model can provide a reasonable fit to the 5 appearance data points, 
even if the fitted function looks nothing like the true oscillation probability. It remains an interesting question, whether such a drastic deviation from the standard behaviour could be tested model-independently by other observations, for instance atmospheric neutrinos.
Notice also the relatively large zero-distance effect for the fit with all baselines,
which show the importance of the near detector measurement in constraining non-standard $P_{\alpha\beta}(L \to 0) \neq \delta_{\alpha\beta}$.
Increasing the error bar of this data point by one order of magnitude to $0.1$
decreases $\chi^2_{\rm min}(0.75~{\rm GeV})$ from 7.40 to 5.05.

The analogue plots for $\delta = 270^\circ$ are shown in Fig.~\ref{fig:FitEnergies270},
where one sees that the shape of the oscillation probability is much more oscillatory-like than for $\delta = 90^\circ$ and the $L$-even fit can quite accurately reproduce the true oscillation probability.
This feature explains the lack of sensitivity of the TV test in the region around the maximally T-violating value $\sin\delta = -1$. From this Figure it becomes also clear that this behaviour does not depend on the specific baselines used here; even many more relatively accurate data points at different baselines would not be sensitive to this case.

To understand why this happens,
one should look into the functional form $f(L)$ of the fitting function and the true probability. We provide here a qualitative discussion. For simplicity we set $\theta_{23} = \pi/4$ and consider oscillations in vacuum; the arguments remain valid also when the matter effect is included. The $3\nu$ appearance probability is then given by
\begin{align}
    P_{3\nu} &\approx 2s_{13}^2 \sin^2\Delta 
    + 2 s_{13}\tilde \alpha \Delta \sin\Delta \cos(\Delta + \delta) 
        + \frac{1}{2} \tilde\alpha^2\Delta^2 \,,
    \label{eq:prob3nu}
\end{align}
where $s_{13} = \sin\theta_{13} \approx 0.15$, $\Delta=\Delta m^2_{31} L / 4E$, $\tilde\alpha = \sin2\theta_{12}\Delta m^2_{21}/\Delta m^2_{31}$ with $|\tilde\alpha| \approx 0.028$. The sign of $\Delta m^2_{31}$ determines the mass ordering. Eq.~\eqref{eq:prob3nu} holds for neutrinos; for antineutrinos the sign in front of $\delta$ should be reversed. Evaluating this expression for $\delta=\pm 90^\circ$ we obtain
\begin{equation} \label{eq:probTmax}
        P_{3\nu}(\delta=\pm 90^\circ) \approx 2 s_{13} \sin^2\Delta
        [s_{13} \mp(\pm) \tilde\alpha \Delta]
        + \frac{1}{2} \tilde\alpha^2\Delta^2 
\end{equation}
for neutrinos (antineutrinos).

In performing the TV test, we are attempting to fit the probability \eqref{eq:probTmax} with an $L$-even function. According to our requirement ($iv$) it will have a similar $\Delta$-dependence as Eq.~\eqref{eq:prob3nu} with $\delta=0$ or $\pi$:
\begin{equation}\label{eq:Peven}
    P_\mathrm{even}(\Delta') \approx a \sin^2\Delta' + b \Delta' \sin 2\Delta' + c {\Delta'}^2 \,,
\end{equation}
where $a, b, c$ are three fit amplitudes with $a,c \ge 0$. Disappearance data require $|\Delta'| \approx |\Delta|$.
The expression in Eq.~\eqref{eq:Peven} is equivalent to our $L$-even fit assuming that zero-distance effects are small,
as it happens with the data we used in the analysis.

Let us focus now on the behaviour at 
the first and second oscillation maxima, i.e.,
$|\Delta| = \pi/2, 3\pi/2$. We can neglect the term proportional to $b$, and Eq.~\eqref{eq:Peven} implies 
\begin{align}\label{eq:prob_cons}
P_{\rm even}(3\pi/2) \ge P_{\rm even}(\pi/2) \,.
\end{align}
Comparing this with Eq.~\eqref{eq:probTmax}, it becomes immediately clear that it is not possible to fit the T-violating probabilities when the second oscillation maximum is suppressed with respect to the first one, i.e., for $\delta=+(-)90^\circ$ for neutrinos (antineutrinos).
For the opposite case, i.e., 
$\delta=-(+)90^\circ$ for neutrinos (antineutrinos), the second oscillation maximum is enhanced and consistent with the condition \eqref{eq:prob_cons}. Indeed, one can show that in this case, choosing three reference values for $\Delta$ close to (but not exactly at) the first and second oscillation maximum and the first oscillation minimum, all three coefficients, $a,b,c$, can be fitted. 

Hence, this simple considerations confirm the behaviour visible in figs.~\ref{fig:FitEnergies} and \ref{fig:FitEnergies270} (which are based on full oscillation probabilities including matter effects):
the suppression (enhancement) of the second oscillation maximum due to the sign of the term proportional to $\sin\delta$
is responsible for the (lack of) sensitivity around $\delta = 90^\circ\, (270^\circ)$ for neutrinos, and the opposite for 
antineutrinos. 
Higher energy bins have no sensitivity to TV due to the fact that there is no longer any data point at the second oscillation maximum.

\begin{figure*}[t]
	\centering
	\begin{tikzpicture}
	    \def\x{0.304\textwidth}
	    \def\y{-4.6}
	    \node at (\x,0){\includegraphics[width=0.365\textwidth]{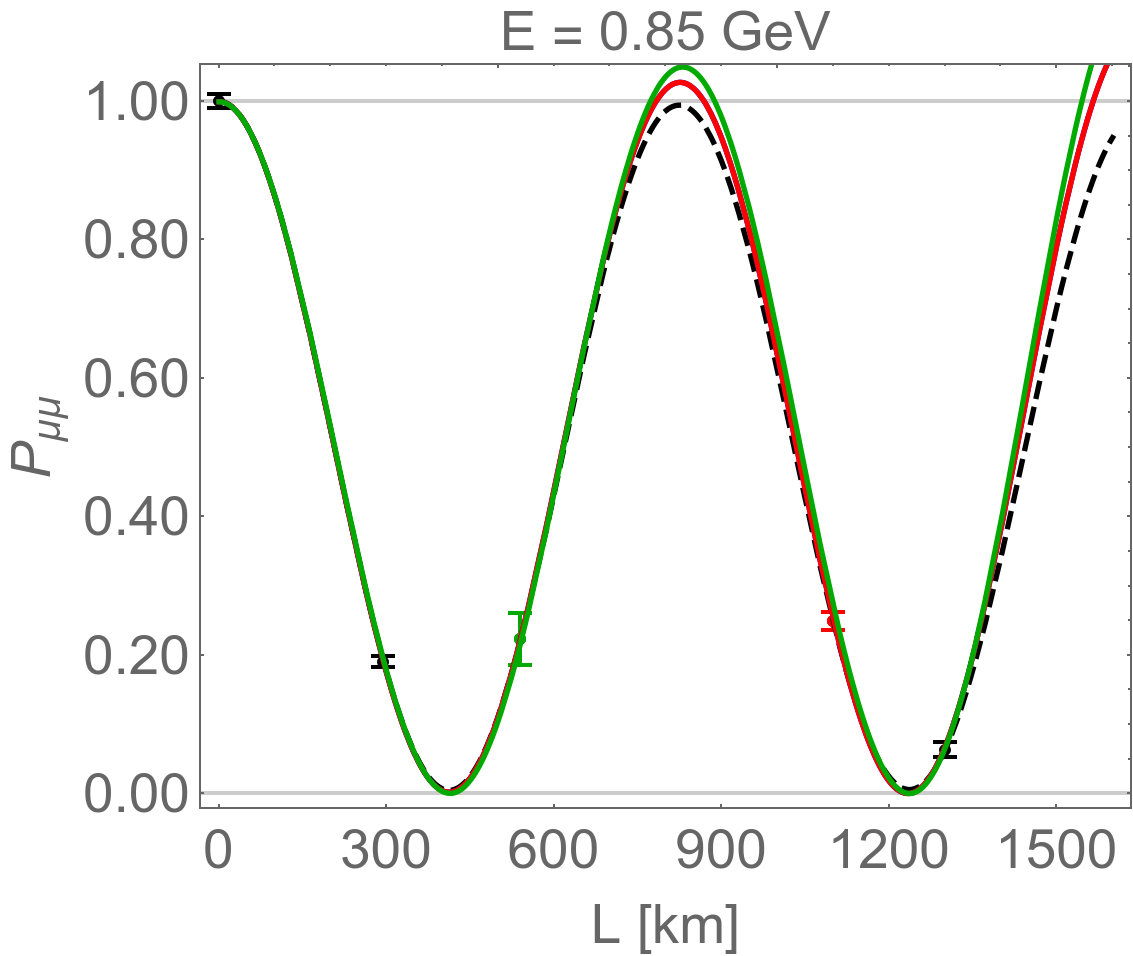}};
	    \node at (0,0){\includegraphics[width=0.365\textwidth]{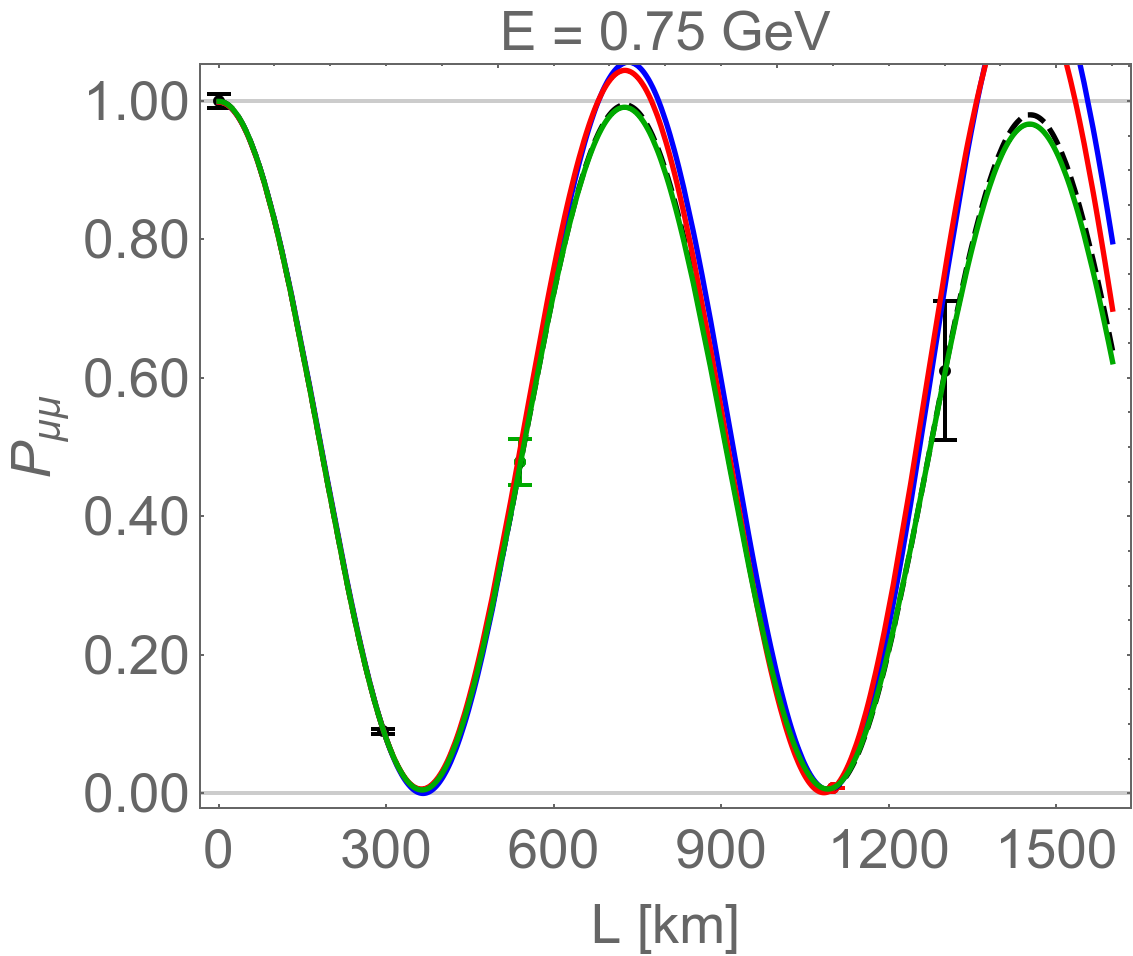}};
	    \node at (-\x,0){\includegraphics[width=0.365\textwidth]{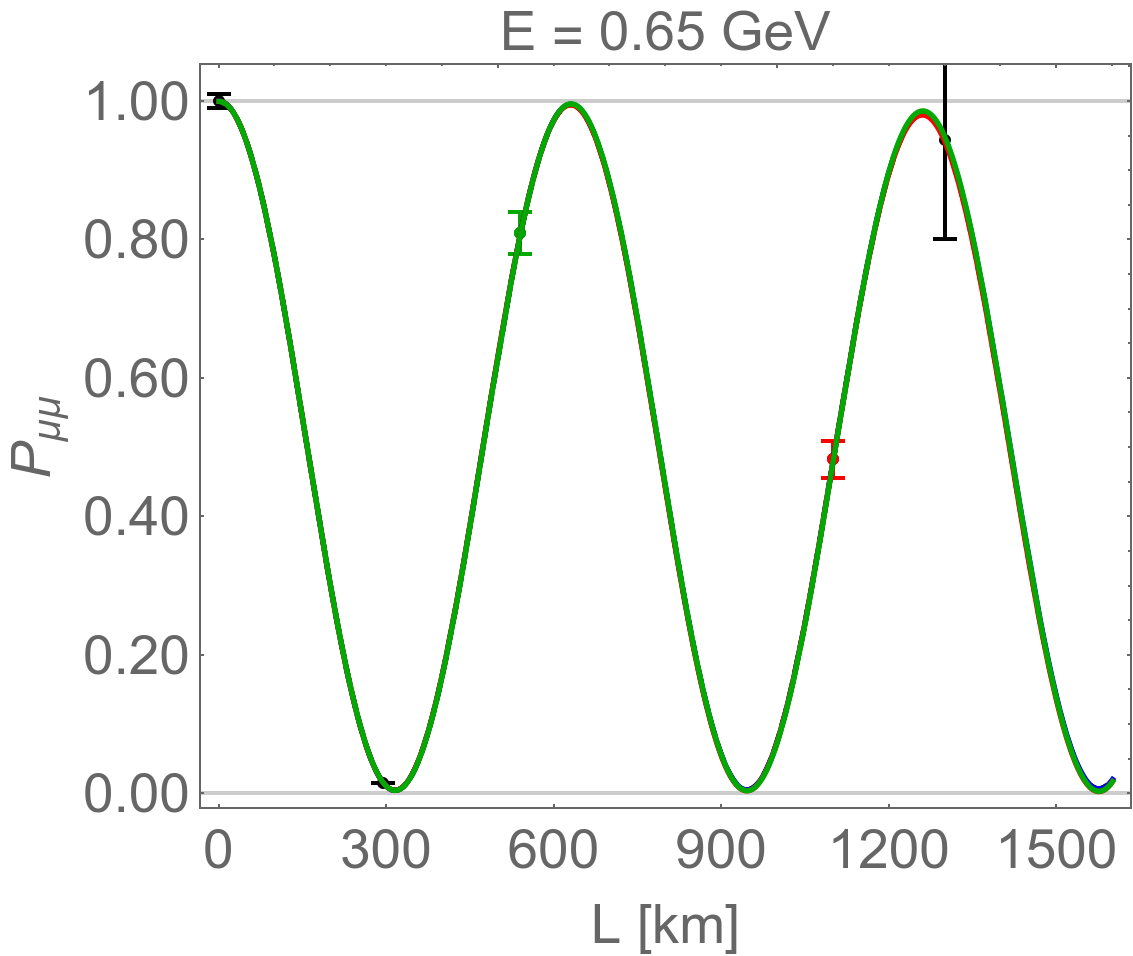}};
	    
	    \begin{scope}[shift={(0,\y)}]
	        \node at (\x,0){\includegraphics[width=0.365\textwidth]{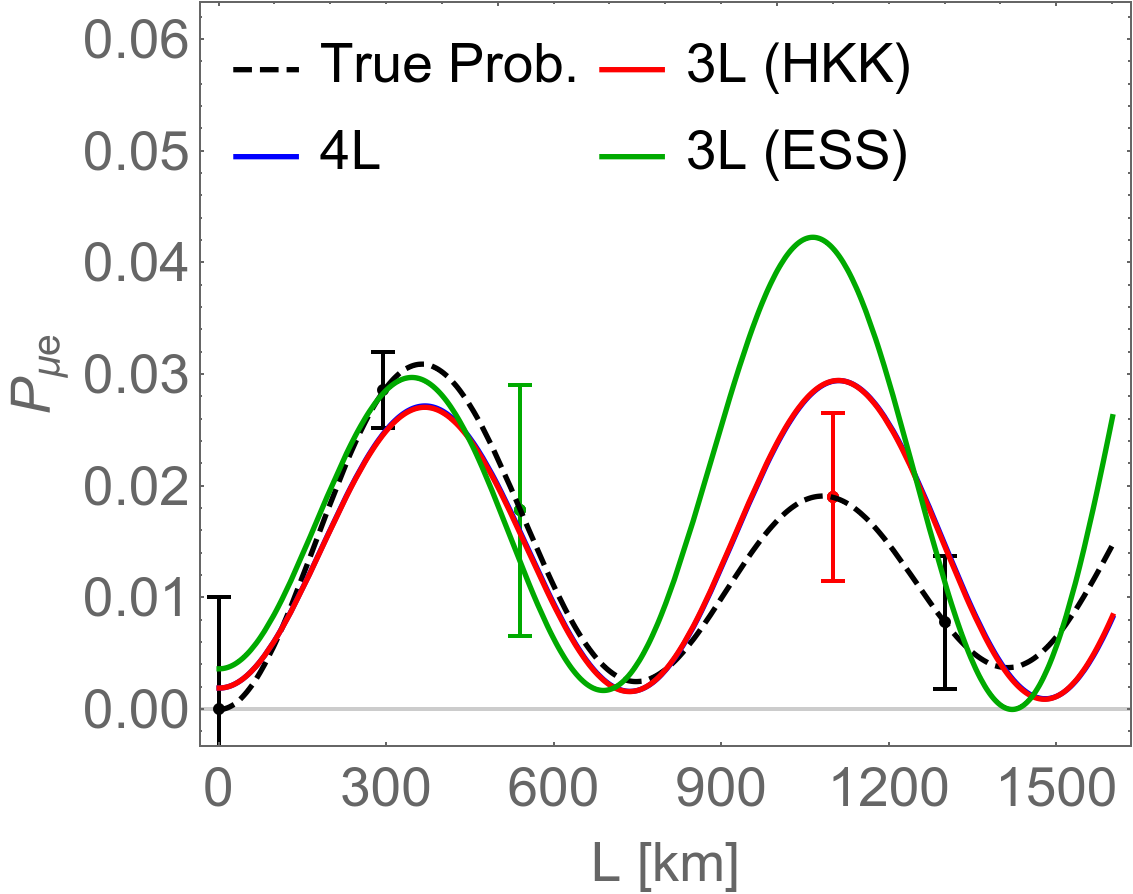}};
	        \node at (0,0){\includegraphics[width=0.365\textwidth]{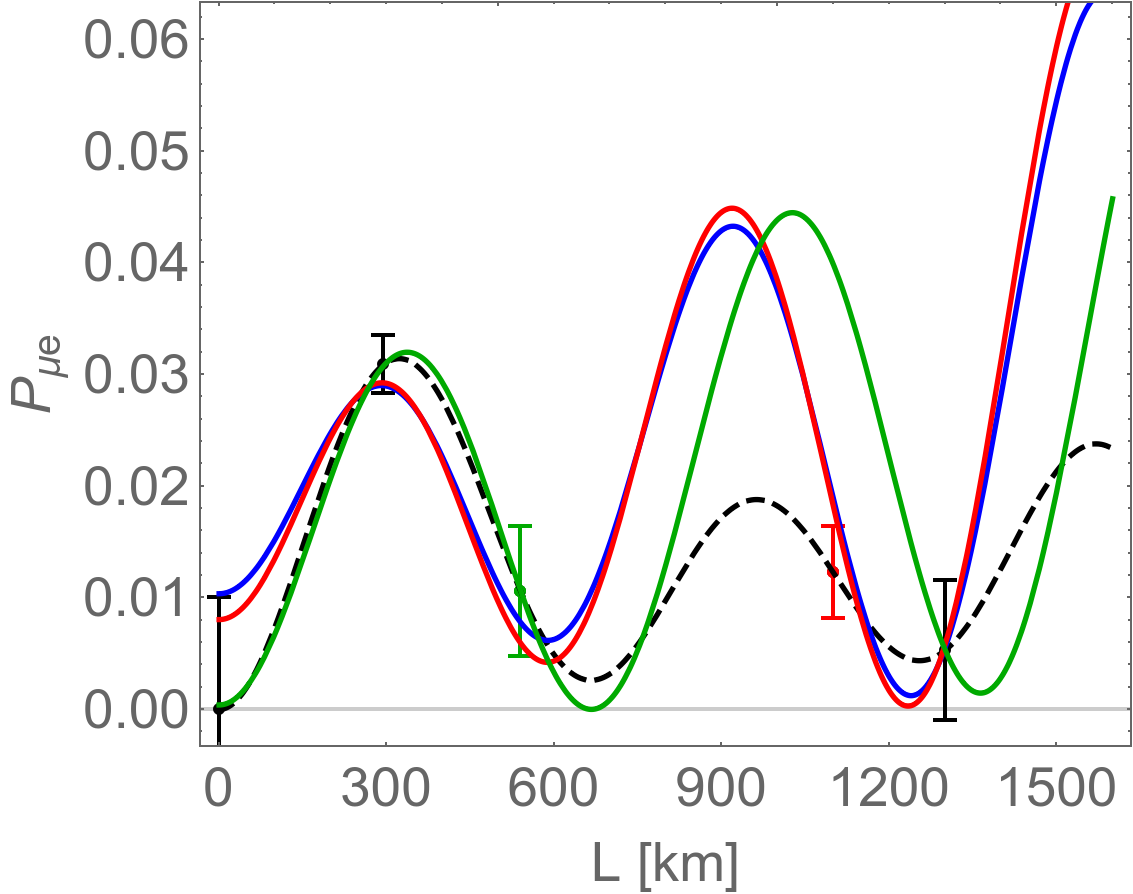}};
	        \node at (-\x,0){\includegraphics[width=0.365\textwidth]{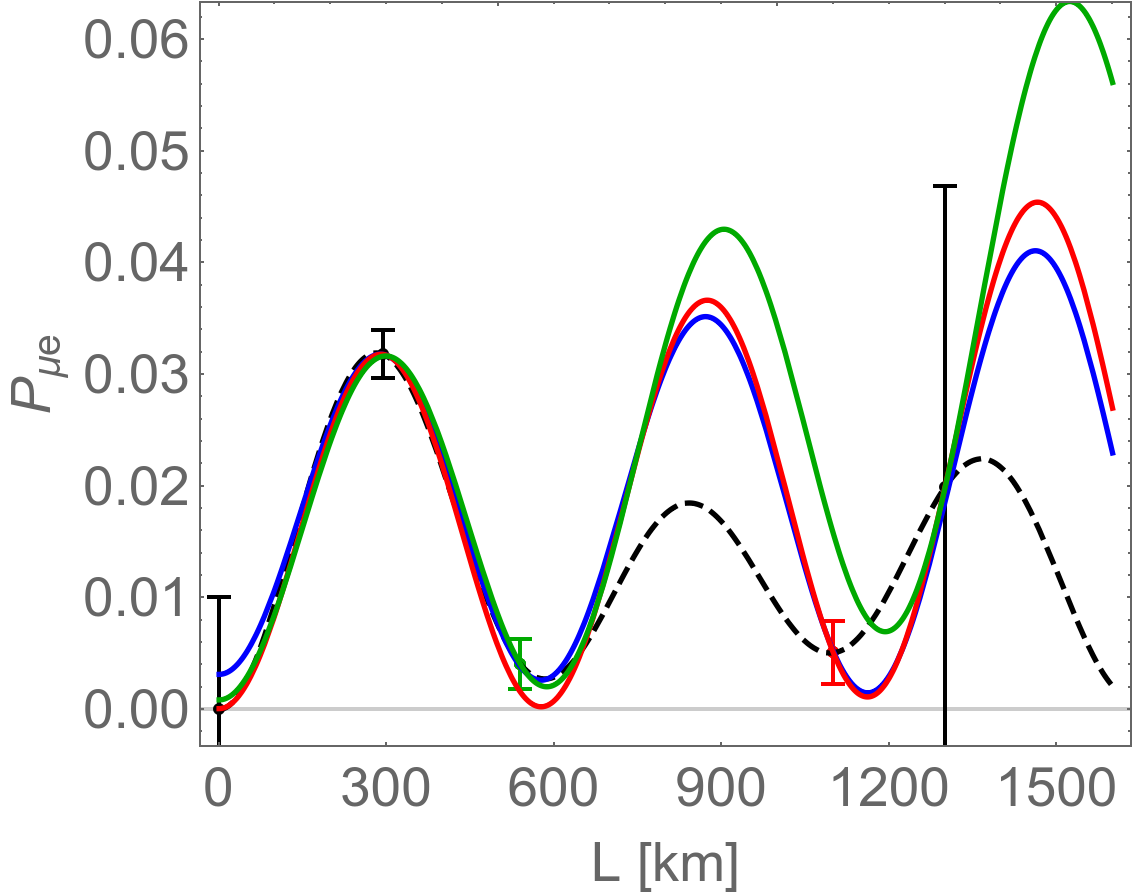}};
	    \end{scope}
	\end{tikzpicture}

	\caption{
	    Same as Fig.~\ref{fig:FitEnergies} for inverted hierarchy.
        }
        \label{fig:FitEnergiesIH}
\end{figure*}

\begin{figure*}[t]
	\centering
	\begin{tikzpicture}
	    \def\x{0.304\textwidth}
	    \def\y{-4.6}
	    \node at (\x,0){\includegraphics[width=0.365\textwidth]{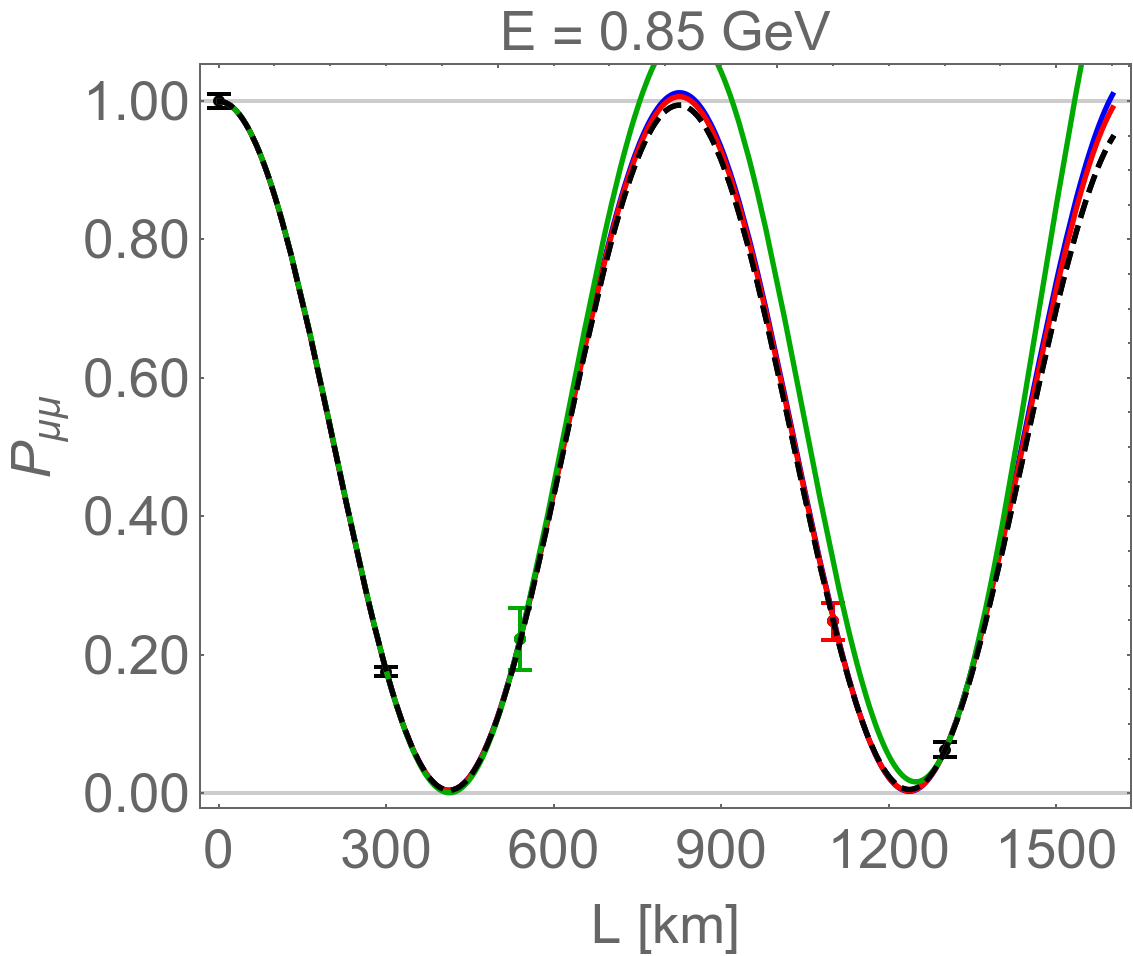}};
	    \node at (0,0){\includegraphics[width=0.365\textwidth]{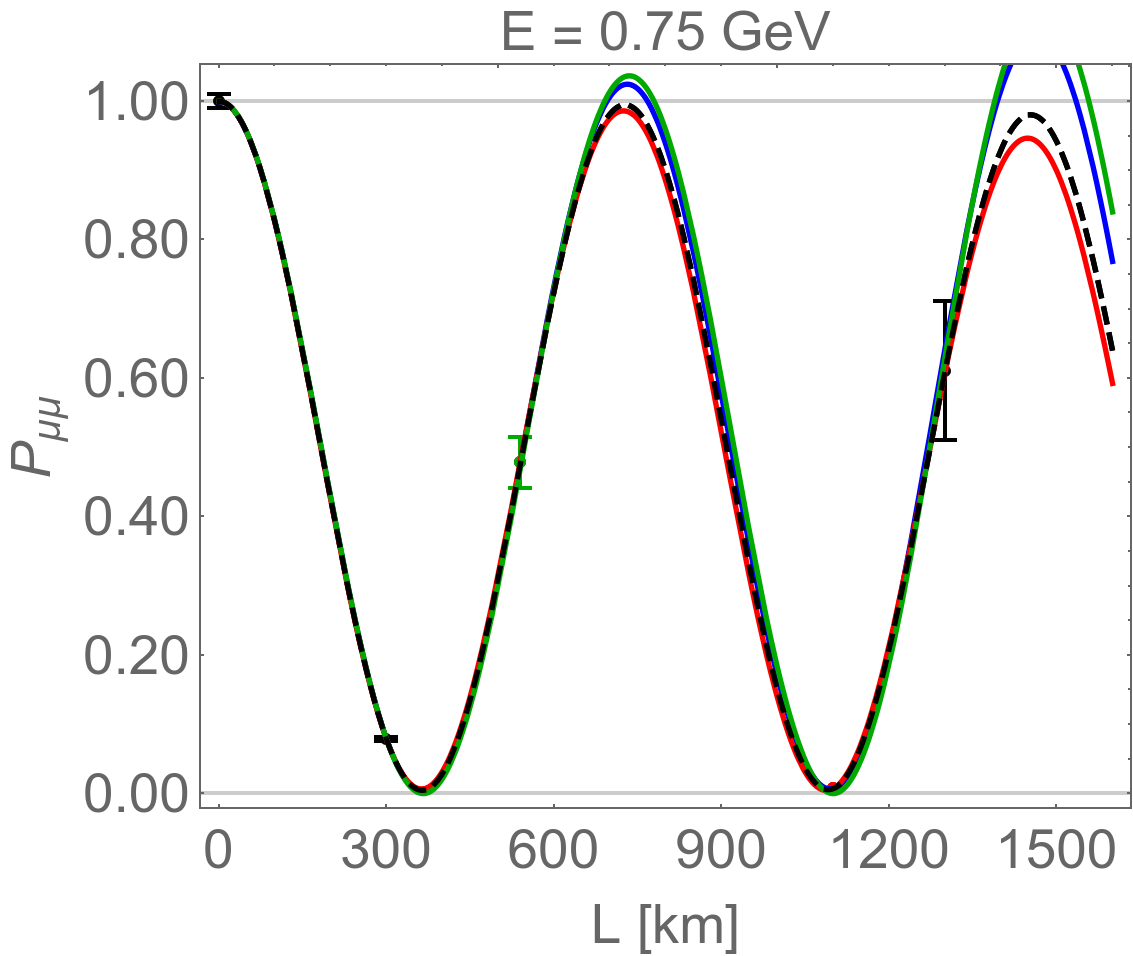}};
	    \node at (-\x,0){\includegraphics[width=0.365\textwidth]{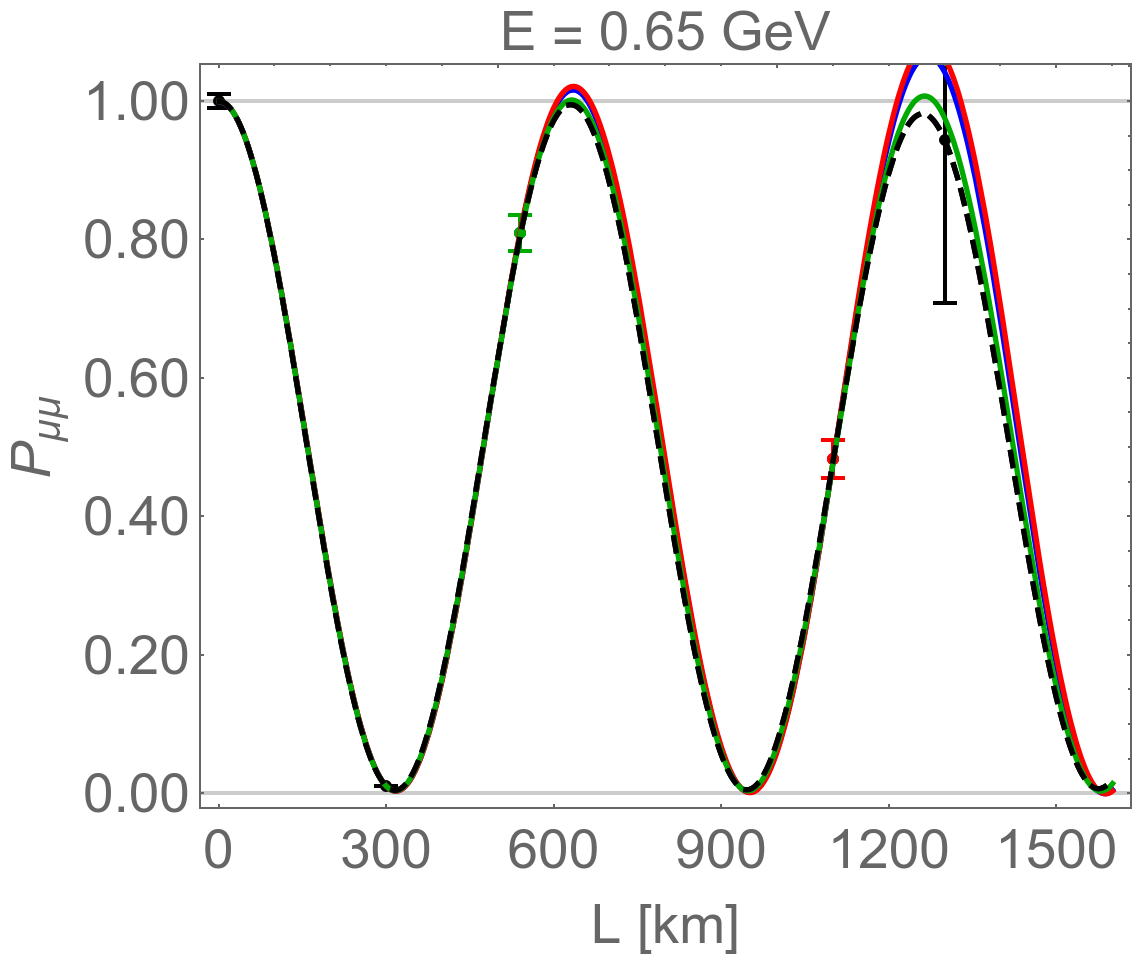}};
	    
	    \begin{scope}[shift={(0,\y)}]
	        \node at (\x,0){\includegraphics[width=0.365\textwidth]{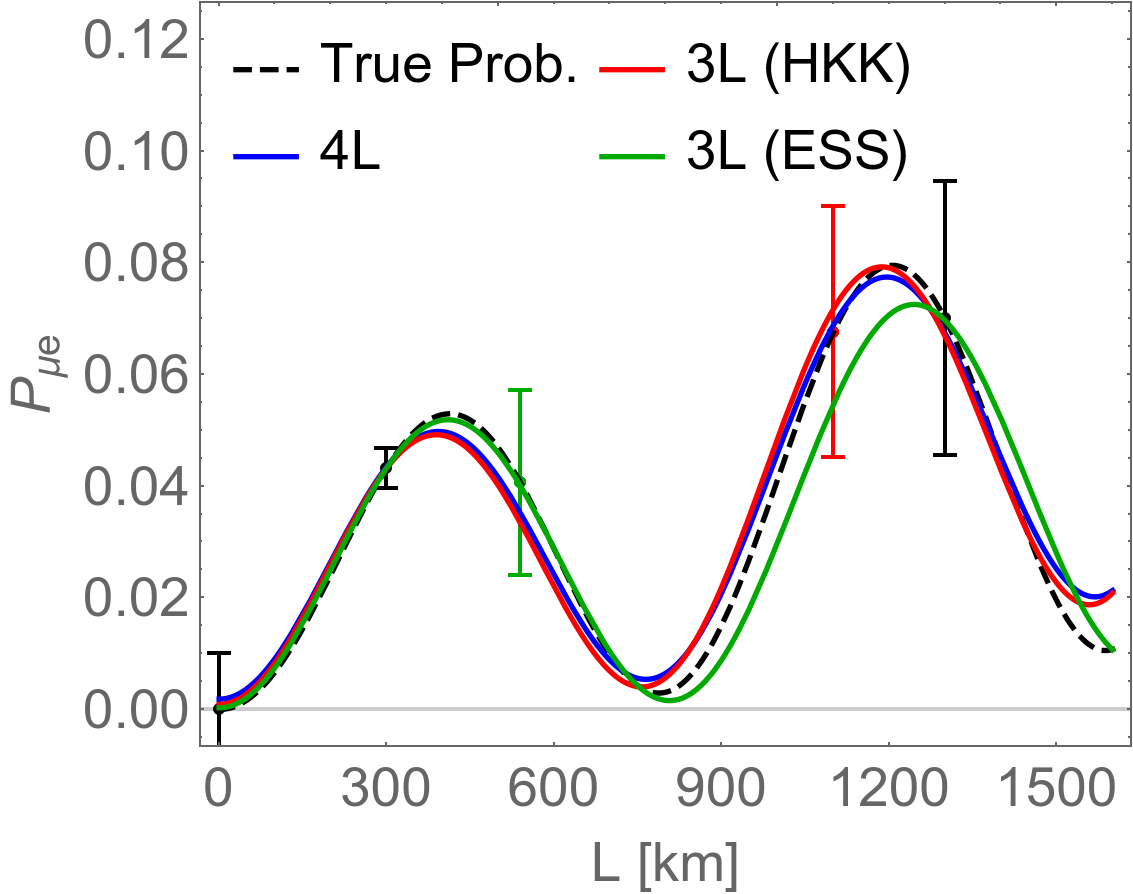}};
	        \node at (0,0){\includegraphics[width=0.365\textwidth]{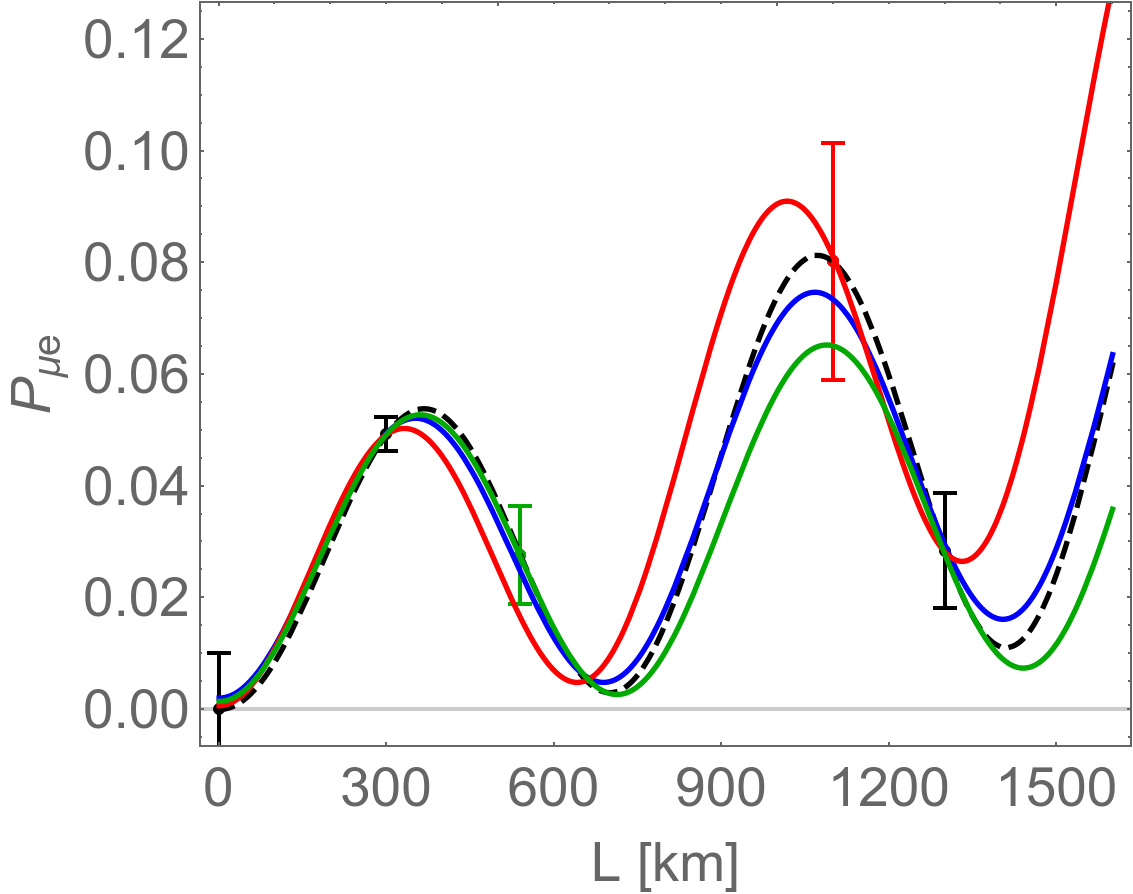}};
	        \node at (-\x,0){\includegraphics[width=0.365\textwidth]{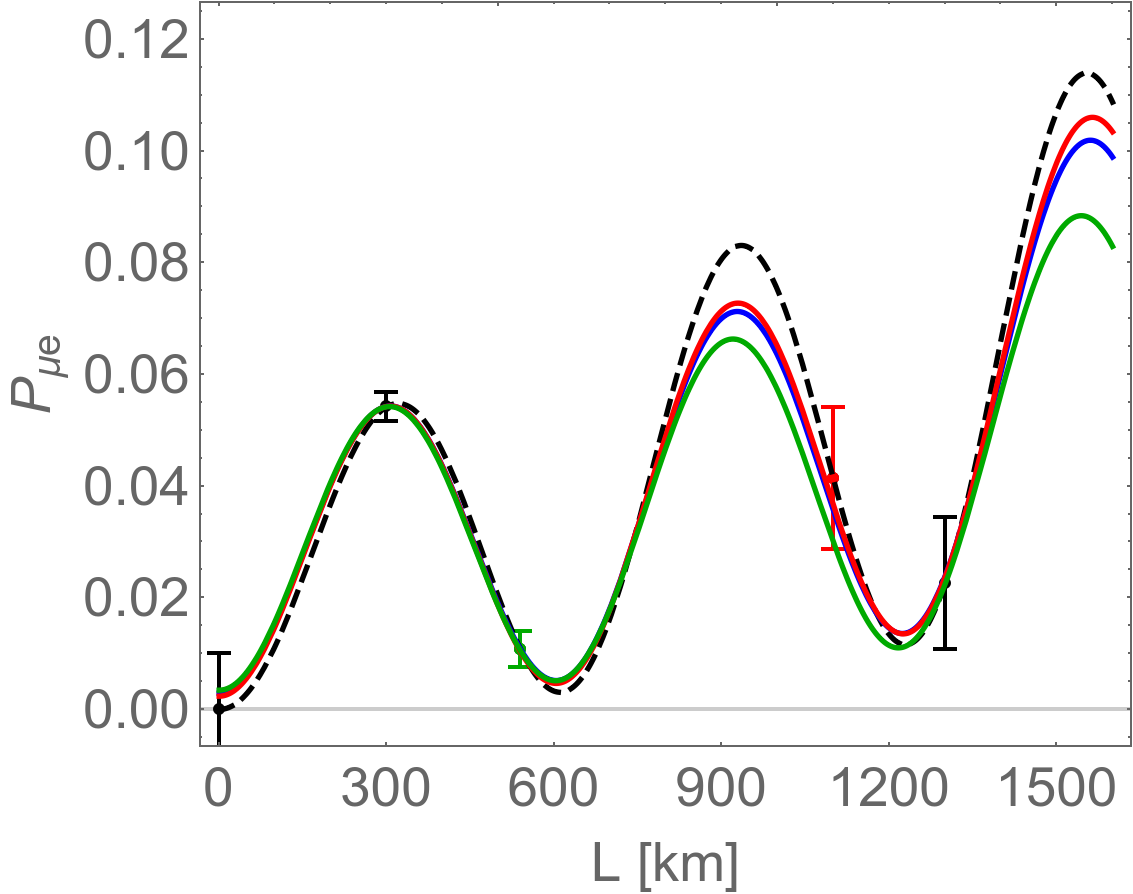}};
	    \end{scope}
	\end{tikzpicture}

	\caption{
	    Same as Fig.~\ref{fig:FitEnergiesIH} for $\delta = 270^\circ$.
        }
        \label{fig:FitEnergies270IH}
\end{figure*}

\begin{table}[tb]
  \caption{
    Same as Tab.~I for inverted hierarchy.
  }
  \label{tab:chi2binsIH}
  \begin{tabular}{ccccc}
  \hline\hline
    $E$     &w/o HKK     &w/o DUNE    &w/o ESS      &all  \\
  \hline
    0.65    &0.01 [0.06] &0.09 [0.70] &0.00 [0.68]  &0.11 [0.70]    \\
    0.75    &0.00 [0.05] &2.46 [4.43] &5.21 [4.36]  &6.01 [4.49]    \\
    0.85    &0.71 [0.59] &1.32 [1.38] &4.87 [2.35]  &4.92 [2.40]    \\
    0.95    &-           &-           &0.71 [0.62]  &-    \\
  \hline
    Tot.    &0.72 [0.71] &3.87 [6.50] &10.78 [7.99] &11.04 [7.59]   \\
  \hline\hline
  \end{tabular}  
\end{table}

Furthermore, it becomes also clear why this behaviour is independent of the mass ordering.
This follows from the observation that the sign of the term $\tilde\alpha \Delta$ in Eq.~\eqref{eq:probTmax} is independent of the mass ordering. We give $\chi^2_\mathrm{min}$ results for inverted ordering in
Tab.~\ref{tab:chi2binsIH}, and figs.~\ref{fig:FitEnergiesIH} and \ref{fig:FitEnergies270IH} show some of the corresponding best-fit curves, which are qualitatively similar to their normal-hierarchy counterparts.
Matter effects induce deviations from the above analytical arguments; however, at the energies under consideration, the matter potential is small enough 
that these deviations are subleading.

\end{document}